# Alterations in Conformations of Poly(3-hexylthiophene) on Au(111) Induced by Annealing

Anmol Arya[1], François Vonau[2], Solomon L. Joseph[1], Thomas Pfohl[1], Silvia Siegenführ[1], Laurent Simon[2*], & Günter Reiter[1*]

[1]Physikalisches Institut, Albert-Ludwigs-Universität Freiburg, Freiburg, Germany
[2]Université de Haute-Alsace, CNRS, Institut de Science des Matériaux de Mulhouse, Mulhouse, France

## Abstract

Employing high-vacuum electrospray deposition and scanning tunneling microscopy, we investigated how individual poly(3-hexylthiophene) (P3HT) chains navigated on the periodic energy landscape of a reconstructed Au(111) surface. The resulting polymer conformations were governed by the interplay between the periodically corrugated substrate, in particular the depth and regularity of the modulated surface potential, and thermal energy. On a regularly reconstructed surface, annealing at 100 °C provided sufficient energy for chain segments to overcome energy barriers of the corrugated surface potential landscape, allowing monomers along the chain to experience a strong thermodynamic driving force toward the low-energy valleys on the surface. The adsorbed polymers adopted a state where the polymer conformations were replicating the herringbone pattern. By contrast, on an irregularly reconstructed surface, the correspondingly disordered potential landscape yielded a diverse mix of coiled polymer chains performing a two-dimensional random walk and collapsed chains located in troughs of the energy landscape. Intriguingly, annealing at 200 °C forced polymers to form clusters of many chains. Our results establish that thermal energy and substrate topography represent control parameters for altering polymer conformations, providing a mechanistic framework for rationally designing polymer nanostructures at the molecular level.

# Introduction

The drive towards miniaturization in organic electronics and nanotechnology has intensified the need to understand and control the organization of conjugated polymers at a molecular scale [1–3]. Nanoscale morphology and supramolecular assembly of conjugated polymers intrinsically govern the performance of organic electronic devices, from field-effect transistors [4] to photovoltaic cells [5–7]. In such systems, charge transport depends critically on conformational order of polymer backbones and their inter-chain packing arrangements [1–3,8]. Using scanning tunneling microscopy (STM) [9–11], significant advances have been made in visualizing and controlling ordering of small molecules on surfaces. However, due to their conformational flexibility and entropy-driven behavior, ordering of polymer chains involves a more complex physical scenario. Consequently, in-depth studies of self-organization processes of polymers on surfaces remain rare [12].

The large conformational entropy of polymers plays a critical role in generating ordered structures, a feature which distinguishes polymers from small molecules [12–15]. For small or rigid molecules, adsorbates on surfaces are primarily characterized by a balance between molecule–substrate interactions and interactions between the adsorbed molecules, often yielding predictable structures or epitaxial monolayers [9–11]. In contrast, a flexible polymer chain explores and navigates over large regions on a surface, often characterized by a complex energy landscape. There, three competing factors determine the finally established polymer conformations.

1) With the drive to maximize conformational entropy, flexible polymers aim to adopt a random coil conformation [12–14].

2) Polymer–substrate interactions may cause adhesion to the substrate and thereby reduce the conformational entropy, which polymer chains often described as a linear sequence of adsorption blobs [16–19].

3) For $\pi$–conjugated polymers, intra-chain $\pi$-$\pi$ interactions promote folding of the polymer backbone, generating single-chain clusters or aggregates [2,3,20].

For highly flexible conjugated polymers such as poly(3-hexylthiophene) (P3HT), which possess low inter-ring torsional energy barriers of few tens of meV [21–25] along their backbone, extensive sampling of chain conformations is thermally accessible, amplifying the role of conformational entropy in determining structures and morphology. More information on characteristic features of P3HT is presented in Section 1 of Supporting Information.

As an additional contribution to the polymer–substrate interaction, the nanoscale surface topography and the corresponding variation in the surface potential ($U(x,y)$) of the substrate play a pivotal yet often overlooked role in controlling the arrangement of polymer chains on a surface [15,16,19,26]. Crystalline surfaces are not necessarily atomically flat. Often surface reconstructions create topographic structures like periodic corrugations which are directly reflected in variations of the energy landscape of the surface [27–30]. The Au(111) herringbone reconstruction, featuring stacks of alternating fcc and hcp domains [27,31,32], has proved as a model for a reconstructed surface. Details of this surface reconstruction and its preparation are presented in Section 2 of Supporting Information. The pattern of the herringbone reconstruction of the gold surface is characterized by a well-defined length scale ($D$) of the corrugation. By an appropriate thermal treatment, a regular surface reconstruction of Au(111) and the corresponding energy landscape of this surface can be established [27,31,33,34].

Previous investigations of conjugated polymers on surfaces, such as P3HT on Au(001)[35,36] or on an iodized Au(111) surface[33], have provided valuable insight on the arrangement of polymer chains guided by polymer-surface interactions. However, the role of surface corrugation in controlling the conformations of polymer chains was not explored in any depth.

In this work, we investigated, enabled by the provided thermal energy, how a model conjugated polymer explored a periodically corrugated surface potential. Through our experimental results, we linked the conformations of the adsorbed polymers to the periodicity ($D$) and the depth ($\Delta E_c$) of the corrugations of the surface potential. We hypothesize that the transition from a kinetically trapped to an equilibrium "corrugation-following" state occurred when the thermal energy $k_\text{B}T$ was sufficient to overcome local energy barriers that prevented chain rearrangement at room temperature. Once these kinetic barriers were surmounted, each monomer along the polymer chain experienced the strong thermodynamic driving force caused by the difference in the corrugation energy $\Delta E_c$ which pulled all monomers of a polymer chain toward the nearest valley of minimum potential.

We addressed our hypothesis by combining high-vacuum electrospray deposition (HV-ESD) of P3HT with variable-temperature STM on Au(111) surfaces. We varied the degree of order of the surface reconstruction (regular vs. irregular) and applied different amounts of thermal energy in a post-deposition annealing procedure. Annealing provided sufficient thermal energy to overcome energy barriers for local rearrangements of the backbone [37,38], enabling polymers to explore the surface and to migrate to the low-energy valleys of the corrugated surface.

## Experimental Details

In this study, we have used regioregular P3HT of weight-average molecular weight between 20,000 and 45,000 g/mol obtained from Sigma-Aldrich (product no. 900563). The employed solvents, toluene and methanol, were obtained from Fisher Chemical (product no. $108-88-3$, with a purity of $< 99.8\,\%$) and Chemsolute (product no. 1459.1000, with a purity of 99.8% and with 0.05 % water), respectively. As the substrate, a single crystal of gold (MaTecK GmbH, Jülich, Germany) was used, which was cleaned under ultra-high vacuum (UHV) conditions through two successive cycles of $Ar^+$-sputtering for 30 minutes at a base pressure of $3.1 \times 10^{-6}$ mbar. After each sputtering step, the crystal was annealed at 500 °C for *ca.* $30-45$ minutes and subsequently quenched to room temperature. Within the first 24 hours after completing the cleaning procedure, STM images revealed regions containing several small atomically flat terraces, indicating that the Au(111) surface had reformed broad (111) terraces but had not yet attained its fully reconstructed state. In other regions, which were imaged during the same time window, small three-dimensional Au islands were observed, reflecting an anisotropic step and edge energy landscape on Au(111). STM images taken after storing the cleaned gold surface for 24 hours at room temperature showed an irregular and locally disordered herringbone pattern, characterized by discontinuous soliton walls and non-uniform elbow spacing, indicating that the underlying threading dislocations have not yet fully coarsened. After storing the gold crystal approximately 64 hours at room temperature, its surface evolved toward the thermodynamically favored state and exhibited a well-ordered, periodic herringbone reconstruction with regular elbow spacing and continuous soliton walls across the entire field of view.

A fresh solution of regioregular poly(3-hexylthiophene) (P3HT) was prepared at a concentration of $1 \times 10^{-5}$ mg/mL using a $6:1$ (volume/volume) mixture of toluene and methanol as solvent. Immediately after preparation, we wrapped the vial containing the solution in aluminum foil to prevent any influence of light, i.e., by absorbed photons. Polymers were deposited via electrospraying for 5 minutes (MolecularSpray Ltd, Nottingham, UK). The P3HT solution was delivered at a flow rate of 0.5 mL/h using a syringe pump with an applied bias voltage of 1.9 kV. We maintained the chamber pressure at $2.8 \times 10^{-7}$ mbar before deposition, which increased to $1 \times 10^{-5}$ mbar during spraying.

We observed that the as-deposited polymer chains could not be visualized by STM measurements, probably because the chains were not yet strongly adsorbed. Thus, right after polymer deposition, we heated the sample (by direct current heating) at 100 °C for 12 hours under UHV conditions, followed by quenching to room temperature. An additional sample was annealed for 2 hours at 200 °C, followed by quenching to room temperature.

STM was done at room temperature in the constant-current mode using a Pt/Ir tip. The STM images were processed using software packages of Gwyddion 2.63 [39] and WSxM (5.0) [40], binarizing the images using the Otsu thresholding method from ImageJ/Fiji software [41]. On the resulting STM images, we measured the area of each aggregate via the "Analyze Particles" function. Finally, we generated a molecular model of a single P3HT monomer in Avogadro software [42] using the MMFF94 (Merck Molecular Force Field 94) and superimposed the molecular model of monomer on the STM images with LMAPper software [43].

## Results and Discussion

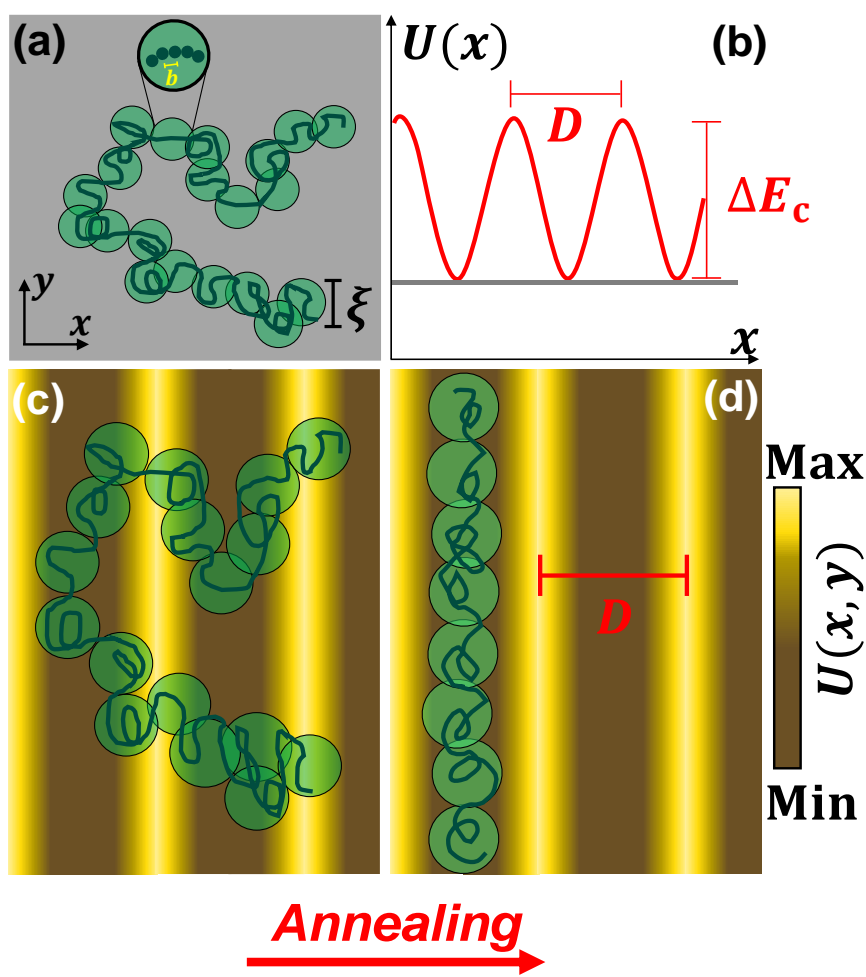

**Figure 1: Schematic representation of a polymer chain confined on a corrugated surface**. (a) As a reference, a polymer chain on a flat equipotential surface with a sequence of adsorption blobs of size $\xi$ and monomer size $b$. (b) 1D representation of the surface potential $U(x)$ (the red curve denotes a corrugated surface potential, while the gray line represents an equipotential surface). $D$ is the corrugation width along the $x$ axis, and $\Delta E_c$ is the energy difference between peak and trough of the surface potential. (c) A polymer chain, represented as a sequence of adsorption blobs, on a periodically corrugated surface potential $U(x,y)$, which only partially follows the corrugations. (d) A polymer chain exclusively adsorbed in a valley of the corrugated energy landscape $U(x,y)$, i.e., following the corrugation completely.

Figure 1 presents a schematic comparison of polymer adsorption on flat surfaces and on periodically corrugated surfaces using the concept of adsorption blobs introduced by de Gennes [44]. This concept provides a framework for understanding how a polymer chain responds to a surface potential by describing the chain as a series of adsorption blobs of size $\xi$. Within each blob the available thermal energy is of order of $k_\mathrm{B}T$ (with $k_\mathrm{B}$ being the Boltzmann constant and $T$ being temperature) which assures that the random arrangement of chain segments within the blob remains unperturbed by their environment. Accordingly, within each blob of size $\xi$, the statistical arrangement of monomers follows a random walk, similar to the one of a free chain in solution. Only on length scales larger than $\xi$, external constraints such as interactions with the substrate become relevant in determining the overall conformation of the macromolecule.

On an ideally flat substrate with a constant surface potential, as indicated in Figure 1 (a), all blobs of the polymer interact uniformly across the entire plane, giving rise to a constant size of adsorption blobs. Thus, on an ideally flat substrate, an adsorbed polymer chain adopts the conformation of a two-dimensional random walk for the sequence of blobs of constant blob size $\xi$.

By contrast, different conformations of polymer chains are expected when polymers adsorb on substrates with a periodically varying surface potential like on the reconstructed Au(111) surface with its regular herringbone pattern. The energy landscape of the corrugated surface illustrated schematically in Figure 1 (b) is characterized by a periodic alternation of low-energy valleys and high-energy peaks. As indicated in Figure 1 (b), the characteristic parameters describing this landscape are the corrugation width $D$, which defines the lateral distance between adjacent peaks, i.e., the width of valleys, and the energy difference $\Delta E_\text{c}$ of these corrugations, which is experienced by a monomer adsorbed in a valley compared to one adsorbed on a peak. For the interaction of a thiophene molecule on a reconstructed gold surface with such a corrugated energy landscape, $\Delta E_\text{c}$ is typically on the order of some tens of meV [45].

Upon initial deposition, polymer chains are often kinetically trapped in non-equilibrium conformations [36] and only adopt an equilibrium conformation after a sufficient duration of an appropriate annealing process. For an energy landscape $U(x,y)$ consisting of linear channels (valleys) periodically spaced at a distance $D$, as depicted schematically in Figure 1 (c), a long polymer chain can potentially span across several periods $D$ of such a corrugated surface. Assuming that on such a corrugated energy landscape an adsorbed polymer chain performs a random walk similar to the one on an ideally flat substrate with a constant surface potential, adsorption blobs would experience varying interaction energies with the substrate, more favorable for positions within valleys and less favorable for positions at peaks. To minimize the free energy of the whole chain, all blobs of a chain should adsorb at positions within valleys (details are shown in Figure S4 of Supporting Information). Thus, the chain will try to adjust its conformation in response to the spatially varying interaction potential, provided that sufficient thermal energy is supplied for the required changes in conformation. Thus, as indicated in Figure 1 (d), we anticipate that the equilibrium conformation of the whole chain is given by a linear string of blobs following the corrugation, i.e., templated by the valley floor given that the blob size $\xi \leq D$.

In the here presented experiments with P3HT deposited onto the herringbone reconstruction of Au(111) we aimed to demonstrate how polymer conformation can be altered through confinement and a suitable templating surface topography.

Figure 2 provides experimental evidence how the regularly corrugated Au(111) surface (denoted as Au(R)) directed the adsorption scenario, resulting in the alignment of electrosprayed P3HT chains. In agreement with the concept of adsorption blobs following the modulations in the surface potential $U(x,y)$, the herringbone reconstruction of Au(111) laterally modulated the distribution of the polymer chains, which all were preferentially localized within the potential minima. However, in order to allow the necessary reorganization of the polymer chain deposited by electrospraying, we had to supply sufficient thermal energy via annealing at 100 °C for approximately 12 hours (STM images before annealing are shown in Figure S6 of Supporting Information). Subsequently, the sample was quenched to room temperature and characterized by STM.

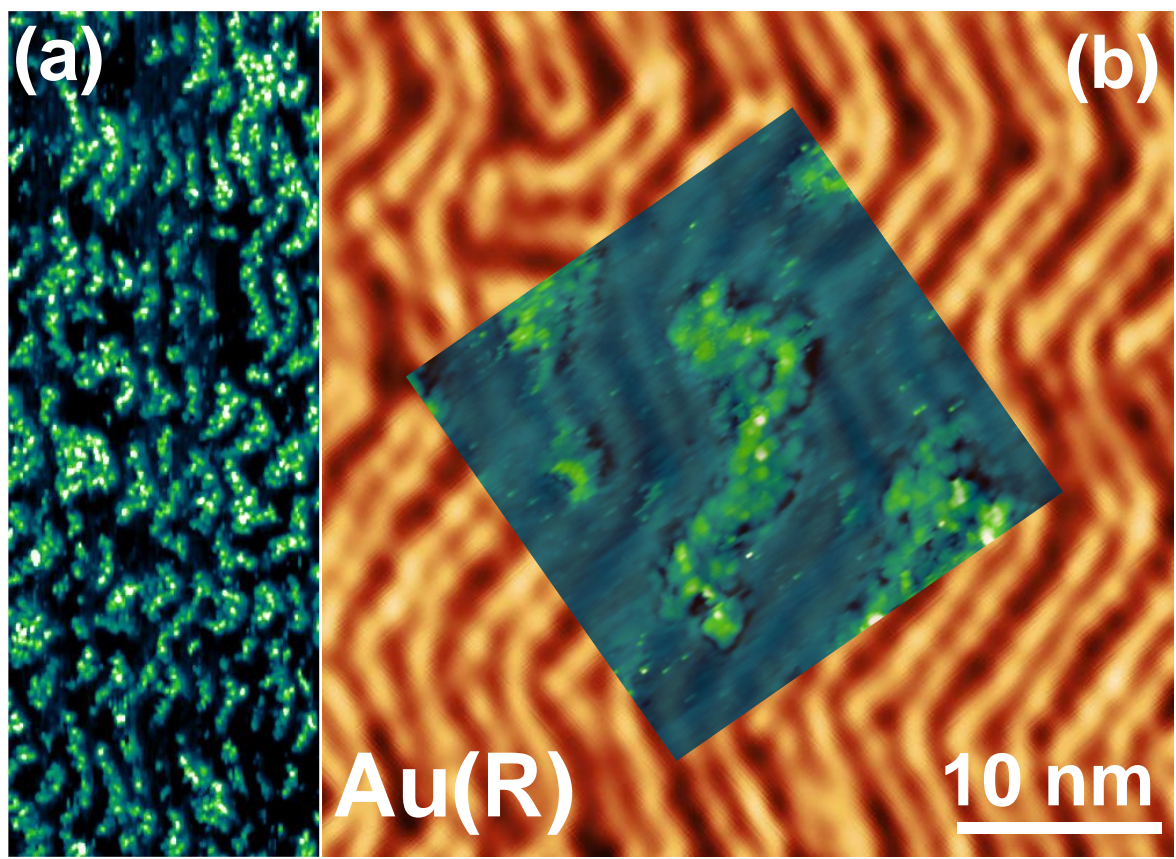

**Figure 2: Influence of Au (111) corrugation on polymer aggregation.** (a) Large-scale STM topography image on Au(R) after electrospray deposition of P3HT, showing that polymer chains followed the herringbone corrugation of Au (111). The size of the image (a) is $62 \times 169$ nm² measured at a setpoint of 10 pA and a bias voltage of – 2.55 V. (b) STM topography image ($42 \times 42$ nm²) of Au (111) before electrospray deposition of P3HT (measured at a setpoint of 1.7 nA and a bias voltage of – 0.2 V), showing the herringbone reconstruction. In the inset, we superimposed a smaller scale topography image ($23 \times 23$ nm²) of P3HT chains deposited on the corrugated surface (measured at a setpoint of 10 pA and a bias voltage of 2.0 V).

The large-scale STM topography image in Figure 2 (a) shows that the deposited polymers followed broadly the surface corrugations of Au(R). Rather than resulting in a random distribution or, alternatively, in an ordered arrangement of $\pi - \pi$ stacked P3HT molecules within a crystalline lamella, the deposited polymers clearly exhibited elongated conformations oriented along the underlying herringbone reconstruction. A periodic separation of molecules by approximately 7 nm was observed across the reconstructed surface, suggesting that the conformations of the chains were templated by the periodic surface potential. In Figure 2 (b), this templating effect is further elucidated. There, STM imaging allowed to resolve simultaneously the deposited polymers and the characteristic herringbone reconstruction, consisting of alternating fcc and hcp domains separated by soliton walls. With the inset in Figure 2 (b), we want to point out that the deposited polymer chains followed the herringbone pattern which is clearly visible in this STM image. The close correspondence between the preferential orientation of the polymer chains and the directions of the underlying soliton walls supports our interpretation that polymer chains were confined within the low-energy regions (the valleys) of the corrugated surface potential. In Figure S8 (c) of the Supporting Information we provide further details.

In order to explore if the regularity of the herringbone pattern was responsible or essential for the templating effect, we deposited P3HT on an imperfectly reconstructed Au(111) surface, which was the result of providing insufficient annealing time for establishing an equilibrated herringbone patterns. Figure 3 demonstrates that the conformations of adsorbed P3HT chains were clearly affected by the missing structural order of the imperfectly reconstructed Au(111) surface.

A typical STM image of the irregular herringbone reconstruction, denoted as Au(IRR), is shown in Figure 3 (a). The surface corrugation lacked a well-defined periodic length scale $D$ and the corresponding surface potential $U(x,y)$ was highly disordered. Thus, as can be deduced from the large-scale STM topography image shown in Figure 3 (b), the deposited polymers also did not show any regular patterns or long range (periodic) order.

This is in stark contrast to the templating effect observed in Figure 2 on regularly reconstructed Au(111), i.e., on Au(R).

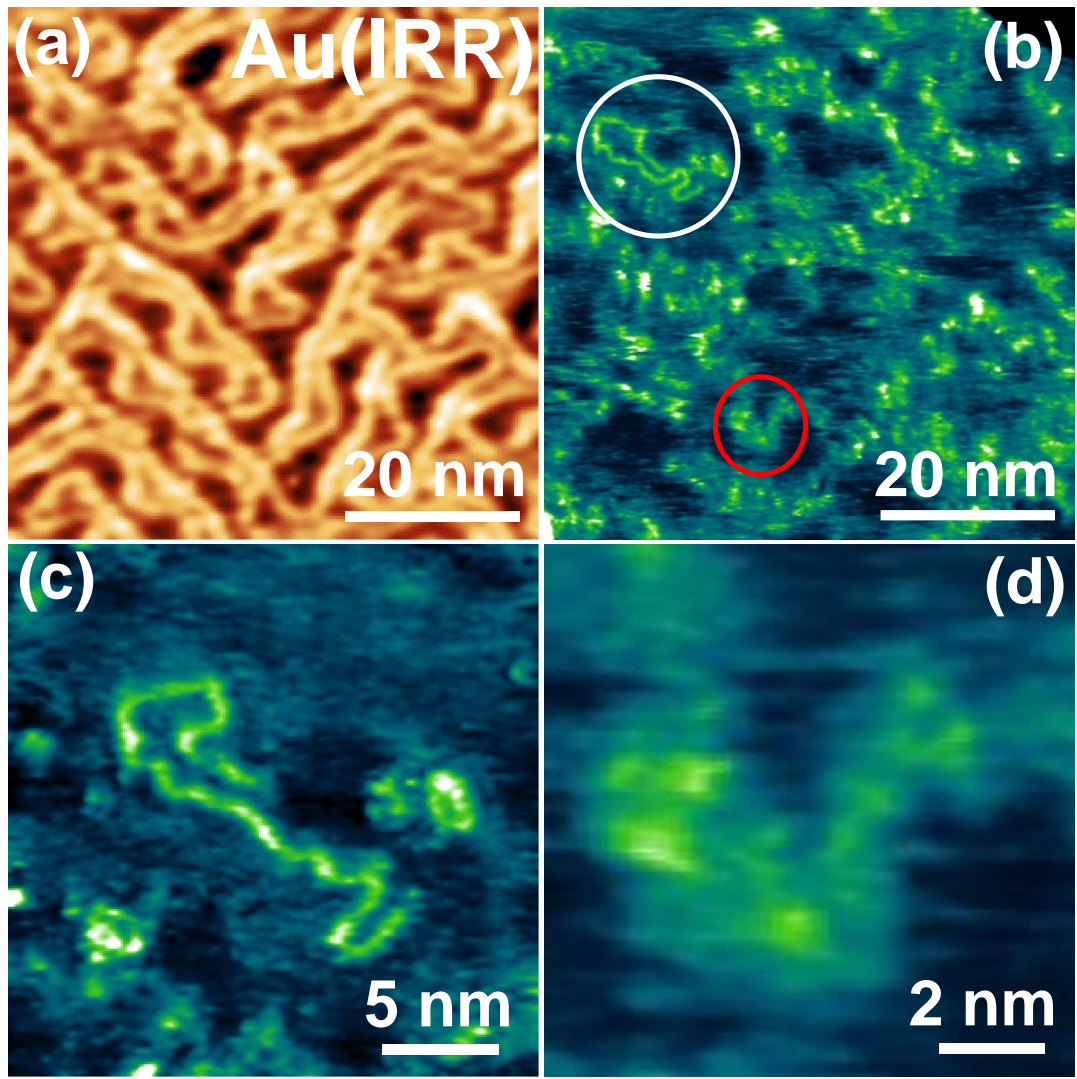

**Figure 3: Influence of irregular corrugations on an Au(111) surface (Au(IRR)) on polymer conformations.** (a) STM topography image of Au (111) with an irregular surface reconstruction before electrospray deposition. (b) Large-scale STM topography image on Au(IRR) after P3HT deposition, showing polymer chains in a 2D random conformations (white circle, see also (c)) and a collapsed state (red circle, see also (d)). The area of image (a) is $60 \times 60$ nm$^2$, measured at bias of $-1.5$ V and a setpoint of 300 pA, (b) of area: $60 \times 60$ nm$^2$ at bias: $-1.51$ V, setpoint: 5 pA, (c) of area $23 \times 23$ nm$^2$ at setpoint of 10 pA, bias voltage of 2.0 V, and (d) of area: $10 \times 10$ nm$^2$ at bias: $-1.51$ V, setpoint: 5 pA.

In contrast to the elongated polymer conformations observed on Au(R), clearly seen on small scale images like the inset in Figure 2 (b), and shown by almost all polymers seen in Figure 2 (a), P3HT chains on Au(IRR) exhibited two distinctly different conformational states. Some chains appeared in a non-collapsed but also non-ordered state. One such example is highlighted by the white circle in Figure 3 (b) and magnified in Figure 3 (c). However, most polymers exhibited more compact, collapsed conformations, as indicated by the example highlighted by the red circle in 3 (b) and magnified in Figure 3 (d).

Figure 3 (c) shows an individual P3HT chain in a non-collapsed conformation. The contour of the chain exhibits a sequence of bends and turns that resembles a two-dimensional random-walk on the surface. In the framework of the blob concept, the different segments within each blob sampled different local minima of the disordered potential, leading to a random, non-templated conformation. Similar images are shown in Figure S10 of the Supporting Information where we also superimposed a model of the polymer chain.

In the small-size image of Figure 3 (d), we show the collapsed state observed for the majority of deposited polymers. Here, probably representing a single P3HT chain, the polymer appeared as a compact object, most likely localized within one of the deeper troughs of the irregular surface reconstruction. This arrangement can be rationalized as a chain whose blobs have become trapped within a particularly favorable (deep) local minimum of the disordered surface potential, representing a locally adapted but globally unorganized conformation. A schematic representation of these two scenarios in Figure 3 (c, d) is shown in Figure S11 of Supporting Information.

Taken together, Figure 3 demonstrates that due to the absence of a regular corrugation pattern on Au(IRR), the deposited P3HT molecules did not exhibit the elongated, substrate-templated conformations observed on Au(R). Instead, P3HT chains displayed a coexistence of random-walk-like and collapsed conformations in local energy minima, reflecting adsorption at the various parts of the disordered surface energy landscape.

As detailed in Section 4 of the Supporting Information, the adsorption energy per blob is of the order of $k_\text{B}T$. The energy difference experienced by a monomer adsorbed in a valley compared to one adsorbed on a peak is $\Delta E_\text{c}$, which is of the order of 50 – 100 meV. In general, upon increasing temperature, the size $\xi$ of the blobs increases and the number of blobs per chain decreases. In addition, upon increasing temperature, we expect that the detachment rate of a polymer chain from positions within the valleys increases, thereby facilitating changes in polymer conformations and enabling diffusion of polymer chains on the corrugated energy landscape of the gold surface.

In an attempt to test this expectation, we annealed a freshly prepared sample at 200 °C for approximately 2 hours under ultra-high vacuum, followed by quenching to room temperature. The temperature of 200 °C remained well below the thermal degradation threshold of P3HT and was also below the nominal melting temperature of crystals of P3HT. We note that for all three samples shown in Figures 2, 3, and 4, the number density of polymers deposited by electrospraying was the identical within experimental uncertainty.

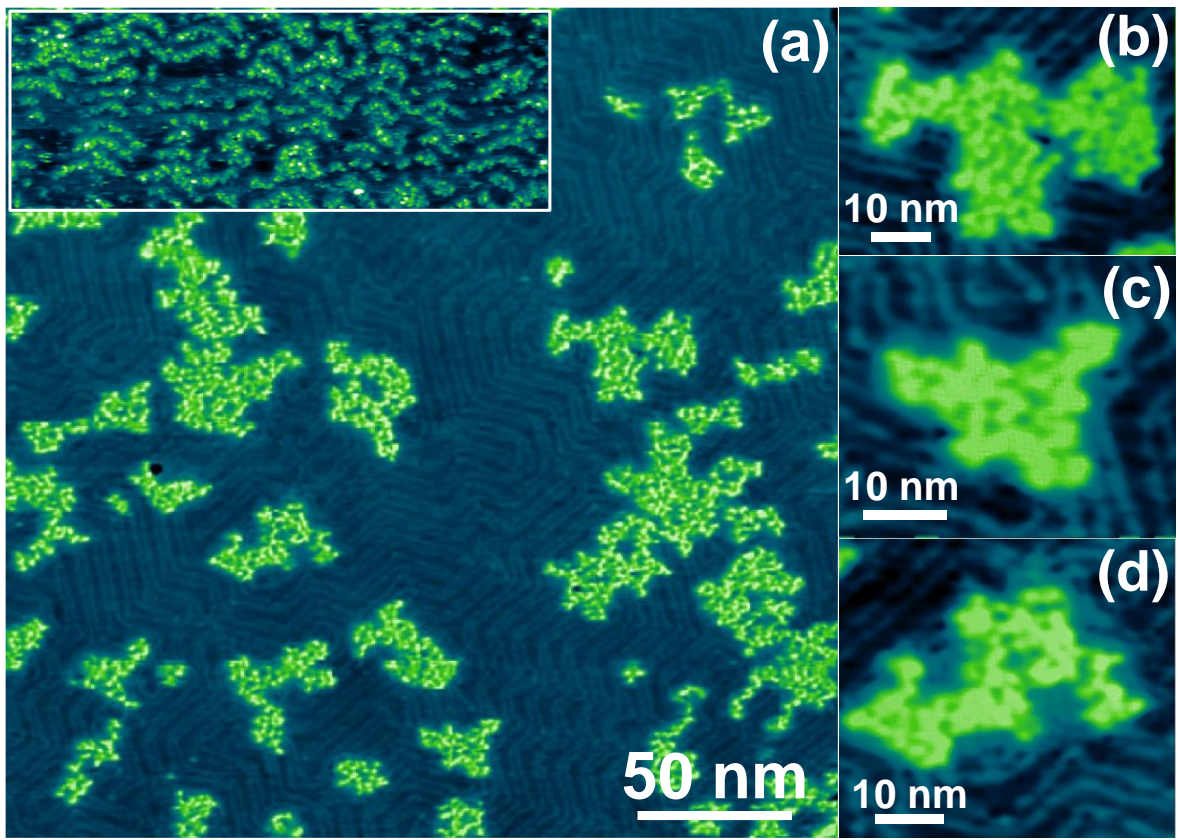

**Figure 4: Thermal annealing resulted in aggregation of multiple P3HT chains**. (a) STM topography image showing clusters consisting of multiple P3HT chains deposited on Au(R) after P3HT and annealed at 200 °C. The scale bar of 50 nm applies also to the inset, which reproduces Figure 2 (a), i.e., a large-scale STM topography image showing P3HT chains after annealing to 100 °C, which followed the herringbone reconstruction of Au(R). (b, c, d) Smaller scale images showing individual clusters of P3HT chains. Image (a) has a size of $269 \times 269$ nm² and was acquired with a setpoint of 120 pA and a bias voltage of − 1.3 V. The inset image covers $169 \times 62$ nm² and was measured at a setpoint of 10 pA and a bias of − 2.55 V. Images (b), (c), and (d) have a size of $57 \times 42$ nm², $40 \times 33$ nm², and $44 \times 50$ nm², respectively, all taken at a setpoint of 120 pA and a bias voltage of − 1.3 V.

We recall that before any annealing the as-deposited polymer chains were not yet strongly adsorbed and could not be visualized by STM measurements. Probably, during electrospraying in vacuum (vacuum is considered as the worst solvent for a polymer chain), chains adopted conformations of a collapsed globule, which enhanced the probability of forming of $\pi - \pi$ stacking interactions within the globule. During annealing at elevated temperatures of 100 °C or at 200 °C, enhanced thermal fluctuations enabled weakening or even separation of $\pi-\pi$ stacking interactions between neighboring thiophene rings. At 100 °C, establishing an enhanced adsorption of a polymer through interactions of several thiophene rings along the backbone of a polymer chain with the gold surface, the resulting chain conformations templated the herringbone reconstruction pattern of the Au(111) surface. However, after annealing a sample at 200 °C for approximately 2 hours, no templating behavior was observed (see Figure 4). As can be deduced from a comparison with the pattern formed after annealing at 100 °C (Figure 2 (a) is re-shown in the inset of Figure 4 (a)), the polymers were much less evenly distributed on the surface. The deposited polymers formed many rather large clusters of several polymer chains, separated by up to 100 nm large regions without any polymer chain. Details about size and distribution of clusters and polymer chains within are provided in Figure S12 and S13 of Supporting Information.

The comparison of results from the two complementary annealing experiments suggests that the loss in conformational entropy and the gain in interaction energy of the $\pi-\pi$ stacks formed within a collapsed chain are of the order of $k_\text{B}T$ at 100 °C. At this annealing temperature, the interaction energy between gold and thiophene rings within the valleys of the herringbone reconstruction is higher than $k_\text{B}T$ at this temperature (i.e., at 100 °C), guiding the adsorbed polymer chains to template the surface corrugations. However, once the thermal energy is higher than the corrugation energy $\Delta E_\text{c}$, which is of the order of 50–100 meV per monomer, the drive toward the valleys of the corrugation is much weaker and chains can diffuse almost freely on the corrugated Au(111) surface, enabling encounters of multiple chains, enhancing the probability of forming multiple $\pi$–$\pi$ stacking interactions between neighboring chains, including the possibility of chain folding and the formation of ordered domains of multiple chains.

## Conclusions

In this study, we have dissected the factors governing alterations in chain conformation and changes in the aggregation behavior of a model conjugated polymer, poly(3-hexylthiophene) (P3HT), on a reconstructed gold surface. By combining controlled electrospray deposition of individual polymer chains at low surface coverage in high-vacuum, appropriate thermal annealing, and detailed characterization via high-resolution scanning tunneling microscopy, we have gained a mechanistic understanding for the adsorption of a conjugated polymer on a corrugated metal substrate.

Our results demonstrated that the nanoscale morphology of a polymer on a substrate is not an intrinsic property of the polymer alone but a tunable outcome dictated by the interplay between the surface potential $U(x,y)$ and the available thermal energy ($k_\text{B}T$). We have shown that on the periodic herringbone pattern of an Au(111) surface, thermal annealing at 100 °C provided the required activation energy for chain segments to overcome energy barriers on the substrate. The corresponding thermodynamic driving force, characterized by the corrugation energy $\Delta E_\text{c}$ of 50–100 meV per monomer, directed all monomers along the polymer chain into the valleys, resulting in a sequential "directed" motion of all polymer chain segments and leading to polymer conformations replicating the corrugated surface potential. Thermal annealing and the corrugated energy landscape of the substrate allowed chains to transition from a state of kinetically trapped conformations inherited from deposition by electrospraying to a state where chain conformations followed the corrugation pattern as a linear sequence of adsorption blobs. The resulting pattern derived from the arrangement of the deposited polymer chains templated by the periodic herringbone pattern.

The decisive role of a well-defined surface potential was confirmed by experiments on an irregularly reconstructed surface. The absence of a periodic template disrupted the directional influence of substrate–polymer interactions, yielding a diverse conformational landscape characterized by a coexistence of two-dimensional random polymer coils and chains collapsed into local potential traps. This control experiment on an imperfectly reconstructed surface highlighted that the surface corrugation acted as a switch, toggling the system between an entropy-dominated regime characterized by random polymer coils and an enthalpy-dominated regime where polymer chains replicated the surface potential. Finally, we have established thermal energy as a powerful handle to drive morphological transitions at elevated temperatures. Annealing at 200 °C, polymer diffusion on the surface increased and allowed neighboring chains to get in close proximity and to form clusters, potentially driven by $\pi - \pi$ stacking among each other.

From an energetic perspective, our observations can be rationalized by comparing the characteristic energy scales of thermal energy responsible for molecular fluctuations with that of surface corrugations and that of $\pi - \pi$ interactions between thiophene rings. At room temperature, thermal energy $k_\mathrm{B} T$ is already comparable with or even larger than the interaction energy of intrachain $\pi - \pi$ stacks of thiophene rings. Thus, such intermolecular interactions are probably too weak to stabilize ordered conformations against thermal fluctuations. At 100 °C, $k_\mathrm{B} T$ becomes comparable to the corrugation energy $\Delta E_\text{c}$ per monomer, enabling chain segments to overcome local barriers of the Au(111) energy landscape and to explore the periodic surface potential. In this regime, the surface corrugation provides the dominant thermodynamic bias, pulling monomers into valleys and enforcing corrugation-following conformations, while intrachain $\pi - \pi$ interactions are still less dominant.

Upon annealing at 200 °C, however, the enhanced mobility allows chains to assemble into multi-chain clusters. Even though each individual $\pi - \pi$ contact is weaker than $k_\text{B}T$, the collective contributions of many such contacts in a cluster becomes comparable to, and can compete with the energy differences of the surface corrugations. A sequence of thiophene rings along the backbone of a polymer chain can interact with a corresponding sequence of thiophene rings along the backbone of a neighboring polymer chain. The simultaneous interaction of multiple thiophene rings may shift the balance from polymer conformations which template the surface corrugations toward clusters of main polymer chains which are stabilized by multiple $\pi - \pi$ interactions between neighboring polymer chains.

Collectively, our work describes a versatile platform and indicates a set of design principles for manipulating polymer nanostructures at the molecular level. By rationally engineering the substrate topography, and thereby the surface potential, and appropriately choosing the thermal processing pathway, one can exert hierarchical control over the conformation of individual polymer chains and their supramolecular aggregation. Within the framework of controlling characteristic energy scales, we may identify routes for tailoring the spatial arrangement of polymer nanostructures, essential for optimizing optoelectronic properties of future organic electronic devices.

**Author Information**

* Corresponding Author

E-mail: guenter.reiter@physik.uni-freiburg.de (G.R).

**Author Contribution**

A.A conceptualized the idea, designed the experiments, prepared samples, and conducted all the experiments. F.V and S.J assisted in STM measurements. S.S and F.V assisted in electrospray deposition. T.P assisted in data analysis. G.R and L.S supervised the overall work. All authors contributed to the analysis of these results. The manuscript was written through the contributions of all authors.

**Conflict of Interest**

The authors declare no competing conflicts.

**Data Availability Statement**

Data for this study are available from the corresponding author upon reasonable request.

**Acknowledgements**

The authors would like to thank Meirui Fu, Da Huang, and Hagar El Mahalawy for their fruitful discussions and suggestions.

**Supporting Information**

The Supporting Information provides additional theoretical background and experimental data supporting this work. Section 1 details P3HT characteristics, including intrachain torsional barriers and interchain $\pi$–$\pi$ interaction energies. Section 2 describes the Au(111) herringbone reconstruction, its dependence on preparation conditions, and the time-dependent evolution used to obtain regular (Au(R)) and irregular (Au(IRR)) substrates. Section 3 quantifies thiophene–Au interactions: absolute adsorption energy and corrugation energy. Section 4 develops the framework connecting polymer conformation to surface templating, including thermodynamic driving forces, kinetic barriers overcome by 100 °C annealing, and the substrate-mediated clustering mechanism at 200 °C. Section 5 presents supplementary STM topography images, monomer-resolved imaging with molecular model validation, and statistical analysis of sizes.

# References

1. Park, K. S., Kwok, J. J., Dilmurat, R., Qu, G., Kafle, P., Luo, X., Jung, S.-H., Olivier, Y., Lee, J.-K., Mei, J., Beljonne, D. & Diao, Y. Tuning conformation, assembly, and charge transport properties of conjugated polymers by printing flow. *Sci. Adv.* **5**, eaaw7757 (2019).

2. Chang, M., Lim, G., Park, B. & Reichmanis, E. Control of Molecular Ordering, Alignment, and Charge Transport in Solution-Processed Conjugated Polymer Thin Films. *Polymers* **9**, 212 (2017).

3. Kukhta, N. A. & Luscombe, C. K. Gaining control over conjugated polymer morphology to improve the performance of organic electronics. *Chem. Commun.* **58**, 6982–6997 (2022).

4. Yang, D. S., Chung, K. & Kim, J. Controlled alignment of polymer chains near the semiconductor-dielectric interface. *Organic Electronics* **76**, 105484 (2020).

5. Luzio, A., Criante, L., D'Innocenzo, V. & Caironi, M. Control of charge transport in a semiconducting copolymer by solvent-induced long-range order. *Sci Rep* **3**, 3425 (2013).

6. Haneef, H. F., Zeidell, A. M. & Jurchescu, O. D. Charge carrier traps in organic semiconductors: a review on the underlying physics and impact on electronic devices. *J. Mater. Chem. C* **8**, 759–787 (2020).

7. Hutsch, S., Panhans, M. & Ortmann, F. Charge carrier mobilities of organic semiconductors: ab initio simulations with mode-specific treatment of molecular vibrations. *npj Comput Mater* **8**, 228 (2022).

8. Van Der Kaap, N. J., Katsouras, I., Asadi, K., Blom, P. W. M., Koster, L. J. A. & De Leeuw, D. M. Charge transport in disordered semiconducting polymers driven by nuclear tunneling. *Phys. Rev. B* **93**, 140206 (2016).

9. De Feyter, S. & De Schryver, F. C. Self-Assembly at the Liquid/Solid Interface: STM Reveals. *J. Phys. Chem. B* **109**, 4290–4302 (2005).

10. Ferreira, Q., Delfino, C. L., Morgado, J. & Alcácer, L. Bottom-Up Self-Assembled Supramolecular Structures Built by STM at the Solid/Liquid Interface. *Materials* **12**, 382 (2019).

11. Mena-Osteritz, E., Urdanpilleta, M., El-Hosseiny, E., Koslowski, B., Ziemann, P. & Bäuerle, P. STM study on the self-assembly of oligothiophene-based organic semiconductors. *Beilstein J. Nanotechnol.* **2**, 802–808 (2011).

12. Möddel, M., Bachmann, M. & Janke, W. Conformational Mechanics of Polymer Adsorption Transitions at Attractive Substrates. *J. Phys. Chem. B* **113**, 3314–3323 (2009).

13. Sung, W., Sung, J. & Lee, S. Effect of surface undulation on polymer adsorption. *Phys. Rev. E* **71**, 031805 (2005).

14. Gin, P., Jiang, N., Liang, C., Taniguchi, T., Akgun, B., Satija, S. K., Endoh, M. K. & Koga, T. Revealed Architectures of Adsorbed Polymer Chains at Solid-Polymer Melt Interfaces. *Phys. Rev. Lett.* **109**, 265501 (2012).

15. Venkatakrishnan, A. & Kuppa, V. K. Polymer adsorption on rough surfaces. *Current Opinion in Chemical Engineering* **19**, 170–177 (2018).

16. Davis, M. J. B., Zuo, B. & Priestley, R. D. Competing polymer–substrate interactions mitigate random copolymer adsorption. *Soft Matter* **14**, 7204–7213 (2018).

17. Netz, R. R. & Andelman, D. Neutral and charged polymers at interfaces. *Physics Reports* **380**, 1–95 (2003).

18. Ge, T. & Rubinstein, M. Strong Selective Adsorption of Polymers. (2016).

19. Edmondson, M. & Saywell, A. Molecular Diffusion and Self-Assembly: Quantifying the Influence of Substrate hcp and fcc Atomic Stacking. *Nano Lett.* **22**, 8210–8215 (2022).

20. Sun, M., Sun, Z., Zheng, Y., Kim, R., Liu, A. L., Richter, L. J., Gilchrist, J. F. & Reichmanis, E. Preprocessing Affords 3D Crystalline Poly(3-hexylthiophene) Structure. *Chem. Mater.* **37**, 2795–2805 (2025).

21. Yan, X., Xiong, M., Deng, X.-Y., Liu, K.-K., Li, J.-T., Wang, X.-Q., Zhang, S., Prine, N., Zhang, Z., Huang, W., Wang, Y., Wang, J.-Y., Gu, X., So, S. K., Zhu, J. & Lei, T. Approaching disorder-tolerant semiconducting polymers. *Nat Commun* **12**, 5723 (2021).

22. Sabury, S., Jones, A. L., Schopp, N., Nanayakkara, S., Chaney, T. P., Coropceanu, V., Marder, S. R., Toney, M. F., Brédas, J.-L., Nguyen, T.-Q. & Reynolds, J. R. Manipulating Backbone Planarity of Ester Functionalized Conjugated Polymer Constitutional Isomer Derivatives Blended with Molecular Acceptors for Controlling Photovoltaic Properties. *Chem. Mater.* **36**, 11656–11668 (2024).

23. Sutton, C., Körzdörfer, T., Gray, M. T., Brunsfeld, M., Parrish, R. M., Sherrill, C. D., Sears, J. S. & Brédas, J.-L. Accurate description of torsion potentials in conjugated polymers using density functionals with reduced self-interaction error. *The Journal of Chemical Physics* **140**, 054310 (2014).

24. Baggioli, A. & Famulari, A. On the inter-ring torsion potential of regioregular P3HT: a first principles reexamination with explicit side chains. *Phys. Chem. Chem. Phys.* **16**, 3983 (2014).

25. Torras, J. Building a Torsional Potential between Thiophene Rings to Illustrate the Basics of Molecular Modeling. *J. Chem. Educ.* **100**, 395–401 (2023).

26. Wang, S., Mayer, A., Dhima, K., Steinberg, C. & Scheer, H.-C. Imprint-induced ordering of crystallizing polymers below stamp protrusions. *Journal of Vacuum Science & Technology B, Nanotechnology and Microelectronics: Materials, Processing, Measurement, and Phenomena* **31**, 06FB06 (2013).

27. Li, P. & Ding, F. Origin of the herringbone reconstruction of Au(111) surface at the atomic scale. *Sci. Adv.* **8**, eabq2900 (2022).

28. Hasegawa, Y. & Avouris, Ph. Manipulation of the Reconstruction of the Au(111) Surface with the STM. *Science* **258**, 1763–1765 (1992).

29. Preetha Genesh, N., Cui, D., Dettmann, D., MacLean, O., Johal, T. K., Lunchev, A. V., Grimsdale, A. C. & Rosei, F. Selective Self-Assembly and Modification of Herringbone Reconstructions at a Solid–Liquid Interface of Au(111). *J. Phys. Chem. Lett.* **14**, 3057–3062 (2023).

30. Seitsonen, A. P. Electronic structure of reconstructed Au(111) studied with density functional theory. *Surface Science* **643**, 150–155 (2016).

31. Hanke, F. & Björk, J. Structure and local reactivity of the Au(111) surface reconstruction. *Phys. Rev. B* **87**, 235422 (2013).

32. Acevedo-Cartagena, D. E., Zhu, J., Trabanino, E., Pentzer, E., Emrick, T., Nonnenmann, S. S., Briseno, A. L. & Hayward, R. C. Selective Nucleation of Poly(3-hexyl thiophene) Nanofibers on Multilayer Graphene Substrates. *ACS Macro Lett.* **4**, 483–487 (2015).

33. Liu, Y.-F., Krug, K. & Lee, Y.-L. Self-organization of two-dimensional poly(3-hexylthiophene) crystals on Au(111) surfaces. *Nanoscale* **5**, 7936 (2013).

34. Bartelt, N. C. & Thürmer, K. Structure and energetics of the elbows in the Au(111) herringbone reconstruction. *Phys. Rev. B* **104**, (2021).

35. Förster, S. & Widdra, W. Structure of single polythiophene molecules on Au(001) prepared by *in situ* UHV electrospray deposition. *The Journal of Chemical Physics* **141**, 054713 (2014).

36. Förster, S., Kohl, E., Ivanov, M., Gross, J., Widdra, W. & Janke, W. Polymer adsorption on reconstructed Au(001): A statistical description of P3HT by scanning tunneling microscopy and coarse-grained Monte Carlo simulations. *The Journal of Chemical Physics* **141**, 164701 (2014).

37. Bhatta, R. S., Yimer, Y. Y., Tsige, M. & Perry, D. S. Conformations and torsional potentials of poly(3-hexylthiophene) oligomers: Density functional calculations up to the dodecamer. *Computational and Theoretical Chemistry* **995**, 36–42 (2012).

38. Kanai, K., Miyazaki, T., Suzuki, H., Inaba, M., Ouchi, Y. & Seki, K. Effect of annealing on the electronic structure of poly(3-hexylthiophene) thin film. *Physical Chemistry Chemical Physics* **12**, 273–282 (2010).

39. Nečas, D. & Klapetek, P. Gwyddion: an open-source software for SPM data analysis. *Open Physics* **10**, 181–188 (2012).

40. Horcas, I., Fernández, R., Gómez-Rodríguez, J. M., Colchero, J., Gómez-Herrero, J. & Baro, A. M. WSXM : A software for scanning probe microscopy and a tool for nanotechnology. *Review of Scientific Instruments* **78**, 013705 (2007).

41. Schindelin, J., Arganda-Carreras, I., Frise, E., Kaynig, V., Longair, M., Pietzsch, T., Preibisch, S., Rueden, C., Saalfeld, S., Schmid, B., Tinevez, J.-Y., White, D. J., Hartenstein, V., Eliceiri, K., Tomancak, P. & Cardona, A. Fiji: an open-source platform for biological-image analysis. *Nat Methods* **9**, 676–682 (2012).

42. Hanwell, M. D., Curtis, D. E., Lonie, D. C., Vandermeersch, T., Zurek, E. & Hutchison, G. R. Avogadro: an advanced semantic chemical editor, visualization, and analysis platform. *J Cheminform* **4**, 17 (2012).

43. Perdigao, L. M. A. L. M. A. LMAPper – Where scanning probe microscopy and molecular visualisation meet.

44. Rubinstein, M. & Colby, R. H. *Polymer Physics*. (Oxford University Press, Oxford, 2023).

45. Takeuchi, N., Chan, C. T. & Ho, K. M. Au(111): A theoretical study of the surface reconstruction and the surface electronic structure. *Phys. Rev. B* **43**, 13899–13906 (1991).

# Supporting Information

# for

# Alterations in Conformations of Poly(3-hexylthiophene) on Au(111) Induced by Annealing

Anmol Arya[1], François Vonau[2], Solomon L. Joseph[1], Thomas Pfohl[1], Silvia Siegenführ[1],

Laurent Simon[2*], & Günter Reiter[1*]

[1]Physikalisches Institut, Albert-Ludwigs-Universität Freiburg, Germany
[2]Université de Haute-Alsace, CNRS, Institut de Science des Matériaux de Mulhouse, France

## Table of Contents

# 1. Poly(3-hexylthiophene)

Poly(3-hexylthiophene) (P3HT) is a conjugated polymer consisting of thiophene rings linked at the 2- and 5-positions, each having an attached hexyl side chain. Single $\sigma$-bonds connecting adjacent thiophene rings - shown by red lines in Figure S1 - allow for torsional rotation. The optoelectronic properties of P3HT are primarily influenced by $\pi - \pi$ stacking of thiophene rings and the planarity of the backbone. In solution, P3HT typically adopts a non-planar, twisted conformation with a dihedral angle of approximately $30° - 40°$ to minimize steric repulsion between the sulfur atom (indicated by yellow spheres in Figure S1) and the hexyl side chains [1–3]. In the solid-state, the packing of P3HT chains is assisted by $\pi - \pi$ stacking or adsorption onto a substrate, which can drive the backbone toward a planar conformation (a torsional angle of 0° or 180°). In both cases, the overlap of $\pi$-orbitals within a single chain or via $\pi - \pi$ stacking across multiple chains augments delocalization of electrons [4–6]. Quantum-chemical calculations and DFT studies on thiophene and substituted polythiophene oligomers show that the torsional barriers between conformers of a chain are on the order of few tens of meV [7–11], while the barrier from a twisted to a planar geometry is typically < 20 meV per inter-ring bond [12]. These low energy barriers imply that torsional movement along the P3HT backbone are readily thermally activated. As a consequence, thermal fluctuations (with an average energy of the order of $k_\mathrm{B}T$, with $k_\mathrm{B}$ being the Boltzmann constant and $T$ being temperature) compete with ordered packing in the solid-state via $\pi - \pi$ stacking and interfacial interactions [13].

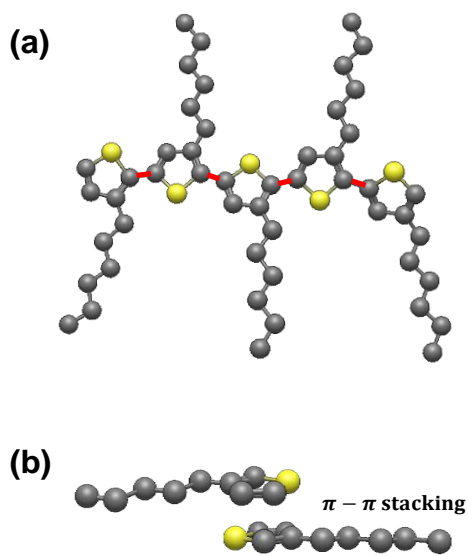

**Figure S1:** (a) Schematic representation of a short regioregular P3HT segment comprising five thiophene repeat units. Carbon atoms are depicted as gray spheres and sulfur atoms as yellow spheres, while red lines indicate the inter-ring $\sigma$-bonds about which torsional rotation occurs. (b) Schematic representation hexyl-thiophene stacked via $\pi - \pi$ interaction[14,15]. For clarity, hydrogen atoms are omitted and the backbone is shown artificially planarized within the plane of the page.

$\pi - \pi$ stacking

One of the governing factors in the arrangement of P3HT at surfaces and in solution is the competition between intrachain forces along the backbone and interchain forces between neighboring chains. This competition determines the conformation of a chain and the final morphology of the polymer film.

Intrachain interactions are dominated by the electronic driving force to delocalize $p$-electrons along the polymer backbone. The energetic gain associated with backbone planarization in thiophene-based oligomers is modest but relevant. Theoretical calculations show that planar and twisted conformers typically differ on the order of 10– 20 meV. Such low torsional energies mean that enhanced planarity, which improves $\pi$-orbital overlap and electron delocalization, can be readily induced by interfacial interactions, promoting the formation of more rigid, rod-like chain segments, except when steric constraints caused by bulky side groups or surface irregularities do not permit such planarization [12].

Intrachain interactions also include $\pi-\pi$ contacts between thiophene units along the same backbone, which can stabilize locally folded or looped conformations and promote single-chain clustering when solvent quality or side-chain design favor intramolecular over intermolecular association [16]. The balance between intrachain $\pi-\pi$ interactions and polymer chain entropy is strongly altered by polymer–substrate interactions, which control whether P3HT adopts extended, clustered, or coil-like conformations at interfaces [17,18].

Furthermore, intermolecular interactions are mediated by co-facial $\pi$–$\pi$ stacking between thiophene rings on adjacent chains, which drives the formation of multichain clusters such as semi-crystalline 2D lamellar structures. Although the interaction energy of an individual $\pi$–$\pi$ contact is only a few $k_\text{B}T$, these interactions are additive along the backbone, so a segment of 10 stacked monomers can be stabilized by an interaction energy on the order of 1000–1500 meV, providing substantial robustness of the cluster at room temperature [19]. In the context of adsorption on a surface, once thermal energy is sufficient for chains to partially decouple from the templating potential of the substrate, cooperative interchain $\pi$–$\pi$ interactions become the dominant driving force for structure formation, promoting rapid coalescence of P3HT chains into multi-chain clusters and crystalline lamellar structures.

## 2. Herringbone reconstruction of the Au (111) surface: Dependence on argon ion sputtering, subsequent thermal annealing, and rapid quenching to room temperature

The crystalline Au(111) surface has been extensively used as a substrate for the deposition of molecules and their characterization at a nanoscale using scanning probe microscopy (SPM) techniques such as scanning tunneling microscopy (STM) and atomic force microscopy (AFM). Atomically flat terraces, high conductivity, and a unique surface reconstruction make gold an ideal substrate for studying molecular ordering, electronic interactions, and interfacial phenomena. Gold (Au), a face-centered cubic (fcc) metal, exhibits a unique reconstruction on its close-packed (111) surface, distinguishing it from other fcc metals. In the bulk, Au(111) lattice planes follow the ABCABC... stacking sequence (Figure S2 (c)), where each atom is coordinated by 12 nearest neighbors. The topmost layer of Au (111) contains 23 atoms per unit length along the [01$\bar{1}$] crystallographic direction, while the bulk lattice aligns with only 22 atoms over the same distance [20]. The reduced coordination of atoms at the surface leads to a higher surface energy. As a thermodynamic adaptation to reduce surface stress and surface energy, the herringbone reconstruction emerges as a characteristic feature of the Au(111) surface. This surface reconstruction is the result of uniaxial contraction, occurring preferentially along one of the three equivalent [1$\bar{1}$0] directions of the substrate. This contraction cannot occur uniformly across the entire layer because the atoms in the top layer remain tethered to the rigid bulk lattice underneath (closed packed {111} planes). As a result, the compressed structure on the surface must accommodate the mismatch with the structure of the bulk, resulting in alternating regions of fcc and hexagonal close-packed (hcp) stacking [21–23]. The fcc regions align with the bulk lattice: surface atoms sit directly above underlying layer in ABCAB arrangement (Figure S2 (c)), preserving the bulk stacking sequence. These regions are energetically favored because they maintain the coordination and bonding geometry of the bulk crystal. By contrast, the hcp regions deviate: surface atoms shift slightly to sit as CBACBA arrangement over the underlying layer (Figure S2 (c)), an arrangement not present in the bulk. This non-ideal stacking introduces strain, making hcp regions less stable. The boundaries between fcc and hcp regions are termed as soliton walls [21–23]. Within these soliton regions, the surface atoms shift above the surface plane by ∼ 0.3 Å. This shift corresponds to a partial dislocation in the hexagonal lattice[23]. The soliton walls act as a mechanism for strain-relief, redistributing the compressive stress from the uniaxial contraction over a finite region. The periodic alternation of fcc and hcp regions along with soliton walls creates a striped pattern with a (22 × √3) unit cell. This uniaxial contraction induces an anisotropic surface stress. To relieve this stress globally, the stripes of soliton periodically bend by ±120° resulting in herringbone reconstruction (Figure S2 (a)). The periodic bends along solitons are localized defects termed as elbows (Figure S2 (a)-EB) [22,23]. Elbow sites are energetically distinct regions of the reconstructed surface. The sudden change in direction of the arrangement of atoms at these sites creates localized strain and lowers the coordination of surface atoms, creating sites of higher surface energy compared to the surrounding fcc, hcp and soliton domains. This higher energy makes elbow sites a preferential region for the adsorption of molecules, nanoparticles, and adatoms [24–29]. During molecular deposition, these sites often act as nucleation centers, guiding the assembly of adsorbates into ordered arrays. Additionally, the threefold symmetry of the substrate permits three equivalent orientations for the uniaxial contraction. This allows the herringbone reconstruction to take three different orientations (at an angle of 120° with respect to each other). Each herringbone reconstruction (from three different directions) consists of parallel rows of alternating fcc and hcp regions. The intersection of these herringbone reconstructions of different orientations results in the formation of curved boundaries with alternating fcc and hcp domains (Figure S2 (b)).

While the thermodynamic equilibrium structure of a clean Au(111) surface is the well-ordered herringbone reconstruction with a characteristic periodicity and minimal defect density, the actual surface observed in experiments depends critically on the thermal history and previous (chemical) treatments of the sample. The reconstruction of the Au(111) surface involves the motion and ordering of gold atoms, as well as the removal of atomic-scale defects which partially require to overcome significant energy barriers. Thus, the equilibrium structure of the Au(111) surface is not easily attainable. Depending on preparation history, non-equilibrium patterns may be observed.

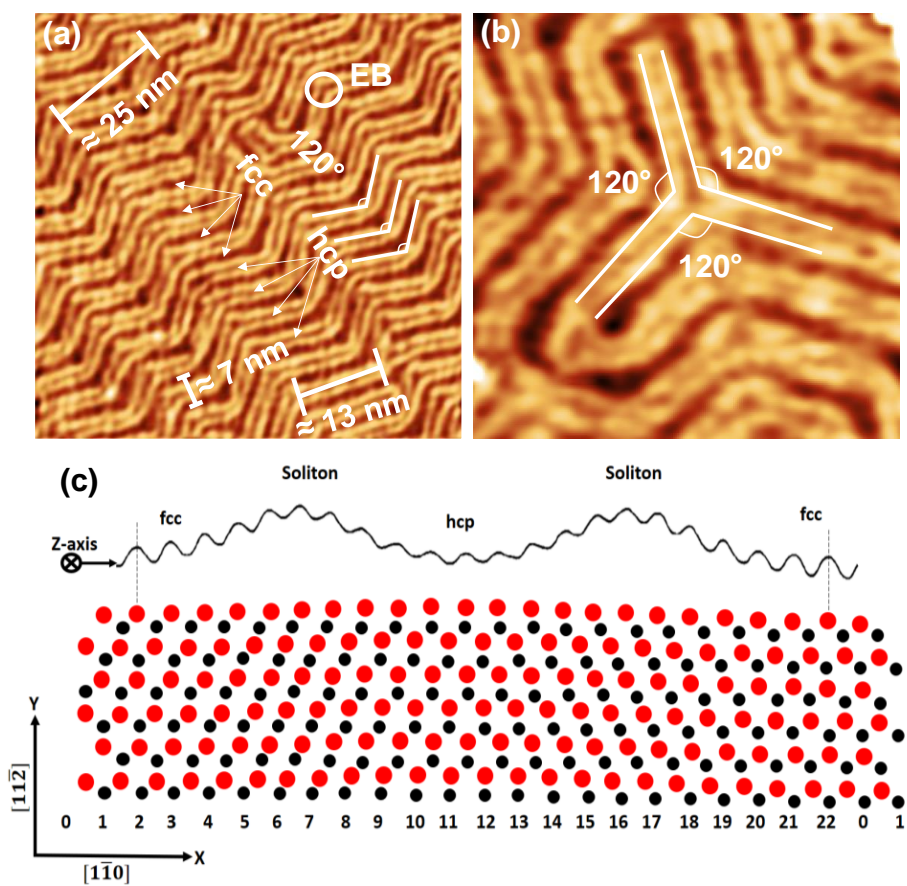

**Figure S2:** (a) STM topography of Au (111) showing zigzag pattern (herringbone reconstruction). The white arrows indicate the fcc and hcp arrangement of atoms. (b) STM topography of an elbow site formed by the intersection of herringbone reconstructions. (c) Atomistic model representation of the $(22 \times \sqrt{3})$ reconstructed Au (111) surface. Red circles correspond to atomic positions within the topmost surface layer,

while black circles represent atoms in the underlying second layer. The soliton walls are represented as topographically elevated domains along the z-axis creating a separation between regions of conventional face-centered cubic (fcc) stacking (ABCABC…) and hexagonal close-packed (hcp) configurations (CBCABC…). Area of the image (a) is 60 × 60 nm$^2$ scanned at bias: −0.200 V, setpoint: 1.7 nA, and for (b) is 7 × 7 nm$^2$ scanned at bias: −1.6 V, setpoint: 100 pA. (c) is Inspired from reference ([20]).

**Argon ion sputtering:** Sample cleaning via Ar$^+$ ion sputtering is the most effective method for removing adsorbates and contaminants from metal surfaces. However, sputtering is an inherently destructive process. Argon sputtering physically removes surface atoms (and sub-surface atoms to shallow depths) through momentum transfer from the incident ions. The energy of typical sputtering processes creates a cascade of displaced atoms, leaving behind a surface that is heavily defective. The sputtering process produces vacancies (missing atoms), adatoms (extra atoms which are not in lattice positions), island defects, and multiple terraces. Most importantly for Au(111) surfaces, sputtering strongly disrupts the Au(111) herringbone reconstruction, producing a highly defective morphology with pits, adatom islands, and step bunching [30]. Thus, recovery of a perfect herringbone reconstruction after sputtering requires post-sputtering annealing at a high temperature. The annealing step must be sufficiently long and the annealing temperature sufficiently high to allow two critical processes to occur:

(1) Vacancies and adatoms must diffuse and either recombine or aggregate, and

(2) Once the surface has regained its flatness (less steps in terraces), the herringbone reconstruction must be nucleated and grow over the whole surface.

The nucleation of the herringbone pattern is a process which involves the collective ordering in many atomic layers. Experimental and theoretical studies have shown that, at room temperature (300 K), the annihilation of defects can take days [23,30].

**Post sputter annealing:** The practical consequence of this surface cleaning procedure is that surfaces prepared by sputtering performed with ions of an energy of 1 − 2 keV followed by insufficient annealing, retain residual disorder, i.e., an incomplete removal of island defects, large number of terraces, and an irregular elbow spacing. Only upon extended high-temperature annealing does the surface heal completely and the herringbone reconstruction may recover a characteristic long-range order within a time period of hours. The annealing temperature represents a fundamental parameter for controlling the degree of long-range order in the herringbone surface reconstruction. At temperatures near room temperature, it can take many hours to days to remove defects. However, at elevated temperatures (several hundred Kelvin above room temperature) diffusion of gold atoms on the surface is much faster, allowing for islands and steps to coarsen significantly [30]. At high annealing temperatures, the thermal energy becomes sufficient to remove defects by overcoming the relevant activation barriers. Ostwald ripening process becomes the dominant mechanism for reducing the density of defects leading to a lowering of the total free energy of the system [30].

**Rapid quenching:** After high-temperature annealing, the sample is cooled to a lower temperature, often room temperature. The protocol of this cooling step is as important as the annealing step itself. When a sample is cooled slowly, the surface can continuously relax and optimize the position of the surface atoms to continuously minimize the free energy. During slow cooling, sufficient time is available for removing dislocations and for adjustments toward a near-equilibrium order. Arriving at room-temperature, the surface may exhibit a regular, long-range-ordered herringbone pattern with minimal frozen-in defects [31].

On the contrary, if the sample is rapidly quenched (cooled suddenly to room temperature), the mobility of atoms on the surface is abruptly lowered and maybe even frozen [30]. Threading dislocations may become immobilized and locked at their current positions on the surface, unable to undergo diffusion and rearrangement processes. The result is a surface with a significant amount of frozen-in disorder: small ordered domains, irregular elbow spacing, high number density of U-shaped domain-wall connectors, and generally poor long-range order [32]. Accordingly, controlled by its preparation history [27,33], an imperfect herringbone reconstruction of the Au(111) surface is not representing a thermodynamically stable structure. However, given enough time is provided, the surface reconstruction will evolve (slowly) toward equilibrium.

In this work, we have used single crystals of Au(111) (MaTecK GmbH, Jülich, Germany) which we cleaned under ultra-high vacuum (UHV) conditions through two successive cycles of Ar$^+$ sputtering for 30 minutes at a base pressure of $3.1 \times 10^{-6}$ mbar, each followed by annealing at $500\,°C$ for $30-45$ minutes followed by quenching to room temperature. Figure S3 presents the time-dependent evolution of the Au(111) surface morphology following the cleaning procedure and quenching to room temperature, highlighting the progressive development of the herringbone reconstruction.

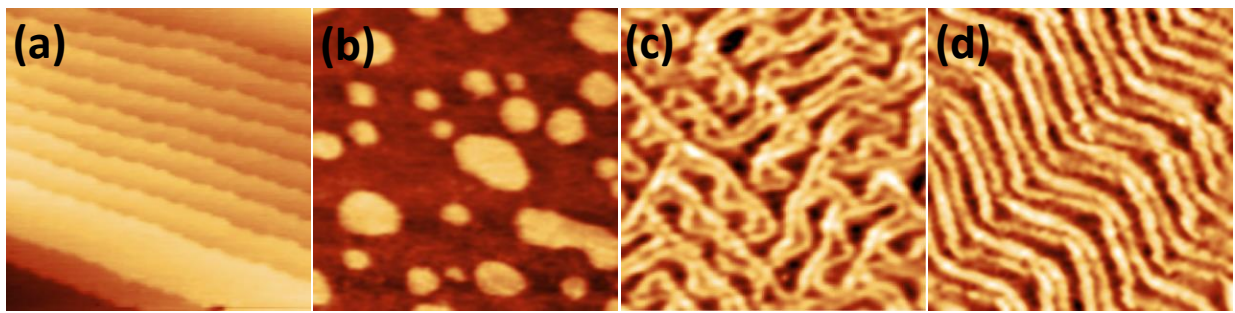

**Figure S3: Time-dependent evolution of the Au(111) herringbone reconstruction following the Ar⁺ ion sputtering cleaning procedure and quenching to room temperature.** (a–c) STM topography images acquired within 24 hours after sputtering and annealing, showing the surface at different locations: (a) atomically flat terraces with limited lateral extension (96 × 96 nm², Bias: −100 mV, Setpoint: 1000 pA); (b) three-dimensional Au island defects aligned along crystallographic directions (25 × 25 nm², Bias: −1.5 V, Setpoint: 1000 pA); (c) an irregular herringbone reconstruction with discontinuous soliton walls and a non-uniform elbow spacing (60 × 60 nm², Bias: −1.5 V, Setpoint: 300 pA). (d) STM topography image acquired 64 hours after cleaning, showing the fully developed, periodic herringbone reconstruction with a regular elbow spacing and continuous soliton walls (20 × 20 nm², Bias: −200 mV, Setpoint: 2300 pA). The progressive evolution from a disordered to an ordered reconstruction demonstrates the slow coarsening kinetics of the Au(111) surface at room temperature.

Within the first 24 hours after preparation, STM images revealed regions containing several atomically flat terraces, with a relatively small lateral extension each, indicating that the surface has largely reformed (111) terraces but has not yet reached its fully reconstructed state. In other areas of the surface during the same time window, small three-dimensional defect islands were observed; consistent with the analysis of Repain et al., these Au islands preferentially aligned along the ⟨100⟩ crystallographic directions [32], reflecting an anisotropic energy landscape on Au(111). In addition, irregular herringbone patterns appeared within the first 24 hours, characterized by discontinuous soliton walls and nonuniform elbow spacing, which indicate that the underlying threading dislocations have not yet fully coarsened. After approximately 64 hours at room temperature, the surface evolved toward a well-ordered, periodic herringbone reconstruction with regular elbow spacing and continuous soliton walls.

# 3. Interactions of aromatic thiophene rings with gold surface atoms

The adsorption of poly(3-hexylthiophene) (P3HT) on Au(111) is governed by two distinct contributions of attractive interactions that are reflected in the total interaction parameter $\mu_o$: the specific localized interaction of the sulfur heteroatom with the gold surface (S-Au interaction) and the interaction of delocalized $\pi$-electrons on the backbone of the conjugated polymer with the metallic bands ($\pi$-Au interaction) [34–37]. The interaction between the sulfur atom of the thiophene ring and the Au(111) surface is the primary anchor for P3HT adsorption. Unlike alkanethiols which undergo oxidative addition to form a strong covalent Au-S thiolate bond (1.8 eV), the sulfur in a thiophene ring does not disturb the aromatic nature of the thiophene ring upon adsorption. The Au-S interaction, involving the shift of electron density from the electrons of sulfur into the empty states of the gold surface, is accompanied by back-donation from the metal $d$-bands into the antibonding orbitals of the thiophene ring [35,38]. Density functional theory (DFT) calculations consistently predict that thiophene rings adsorb preferentially at bridge or hollow sites on the Au(111) lattice, with the sulfur atom positioned closest to the surface to maximize orbital overlap [39]. For a single thiophene molecule, the adsorption energy on Au(111) is calculated to be $0.4 - 0.6$ eV/molecule using van der Waals-corrected DFT functionals (e.g., vdW-DF, PBE+D3) [39]. This energy is significantly weaker than a thiolate bond but stronger than simple physisorption, placing it in the "weak chemisorption" regime. This intermediate strength allows for surface diffusion while maintaining adhesion.

In addition to the localized Au-S anchor, the delocalized $\pi$-electron system of the thiophene backbone of the polymer contributes significantly to the total binding energy through dispersion forces. The conjugated backbone adsorbs with the thiophene rings in a flat-lying (face-on) geometry on Au(111) [40]. This configuration maximizes the contact area between the $\pi$-orbitals and the electrons on the metal surface. DFT studies confirmed that the flat adsorption geometry is energetically favored over tilted or vertical configurations by approximately $0.1 - 0.2$ eV per ring due to this enhanced electronic coupling [39]. The attractive force arises largely from London dispersion interactions (van der Waals forces) between the $\pi-$ electrons of thiophene ring and the high-density electron sea of the gold substrate [41].

# 4. Connecting polymer conformation to surface templating

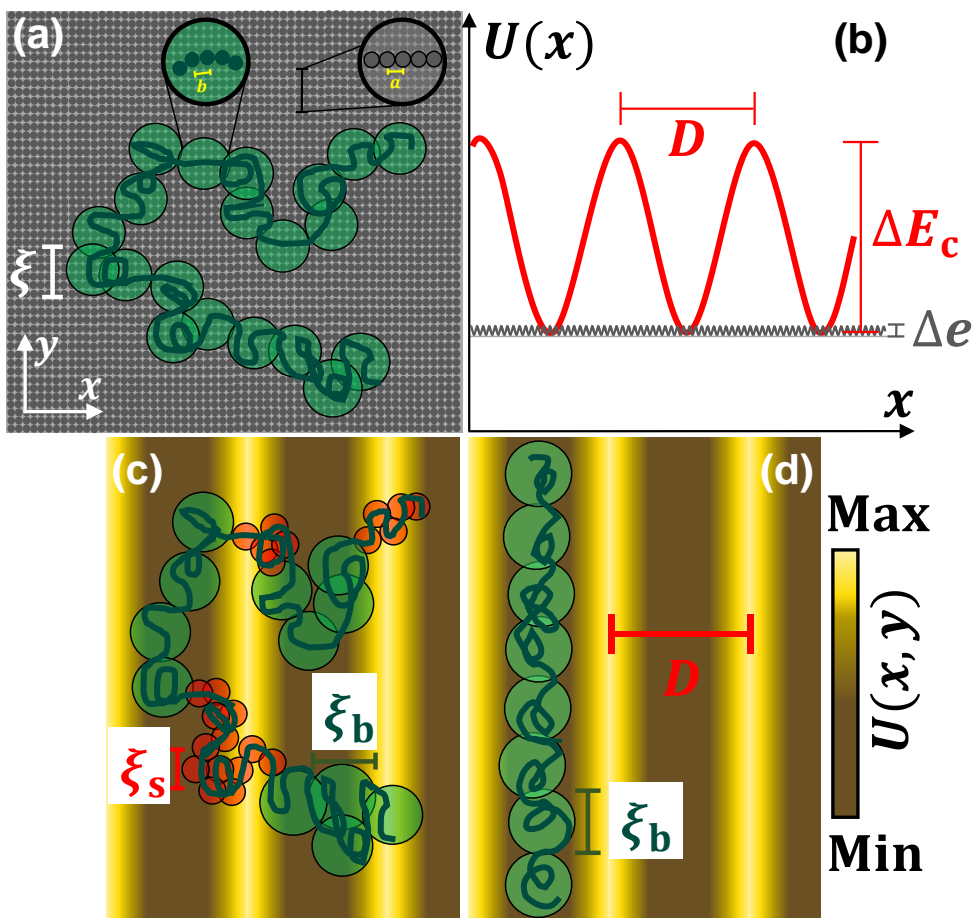

**Figure S4: Schematic comparison of polymer response to surface corrugation.** (a) Surface with a corrugation period $D$ is larger than the size $b$ of a statistical monomer (Kuhn length), which is representing the case of a polymer on an atomically flat surface. Monomers experience an effectively constant surface potential, yielding uniform chain statistics represented by adsorption blobs of equal size $\xi$. The gray dots of size $a$ represent the atomic-scale periodicity. (b) 1-D representation of a corrugated surface potential with a corrugation width $D$ and an energy difference between peak and trough of $\Delta E_c$. On an atomically flat surface, the corresponding atomic-scale periodicity is $a$ and the energy difference between peak and trough is $\Delta e$. Such variations of the surface potential are "invisible" for the polymer given that $a < b < D$. (c) Surface with $D > b$: monomers resolve the variations in the surface potential, leading to position-dependent statistics of the conformations of the polymer chains, with a size of the adsorption blobs varying between $\xi_s$ (smaller on peak) and $\xi_b$ (larger in valley) in the arrangements shown, which is a partially adapted non-equilibrated state. (d) Fully energy-optimized state after thermal annealing: chain confined entirely within valleys, with a constant size $\xi_b \leq D$ of the adsorption blobs.

a) **The concept of "blobs":** The blob concept, introduced by de Gennes (1979)[42], provides a framework for analyzing the statistical mechanics of polymers under a variety of constraints.

For an unperturbed chain in a good solvent, the conformation is that of a self-avoiding random walk. However, when subjected to constraints such as confinement, external fields, or adsorbing surfaces, the statistics of chain conformations are modified. The blob model addresses this by employing a coarse-graining procedure: the chain is divided into a series of "blobs" of a characteristic length (size) $\xi$ (Figure S4 (a)). Within a single blob, the chain segments are unperturbed by the external constraints and exhibit ideal (or self-avoiding) random walk statistics. On length scales greater than $\xi$, however, the cumulative effect of the perturbation becomes significant, dictating the global chain conformation.

In the specific case of a flexible polymer chain confined within a thin film of variable thickness $h(x,y)$, the local blob size is dictated by the local film height, such that $\xi(x,y) \sim h(x,y)$ [42]. This relationship implies that on a non-equipotential surface or, more generally, in a system with a spatially varying potential $U(x,y)$ the blob size $\xi$ cannot remain constant. To minimize the free energy of the overall system, the chain adjusts its local blob size or correlation length. For a chain under the influence of a spatially varying potential $U(x,y)$ as shown in Figure S4 (c, d), the condition for local equilibrium leads to a variation in blob size $\{\xi_s, \xi_b\}$. According to de Gennes' concept of adsorption blobs[42], to minimize the free energy of the system, the size of $\xi$ should change with spatial variations in the interaction potential ($\xi(x,y) \sim -U(x,y)$). A schematic representation of a chain on a non-equipotential surface is shown in Figure S4 (c) with different sizes of $\xi$ (i.e., $\{\xi_s, \xi_b\}$). Upon annealing, the chain is eventually reaching a thermodynamically equilibrium value of $\xi$, as shown in Figure S4 (d).

**(b) The Strong Driving Force for Monomers to Enter Valleys**

(i) **Monomer-surface interaction energy:** Let $\delta$ be the absolute adsorption energy per monomer. For thiophene-based molecules on gold, DFT calculations show that the adsorption energy is substantial, on the order of $0.5 - 1.0$ eV per monomer for the most favorable binding configurations as detailed in Section 3. However, what matters for our analysis is not the absolute adsorption energy, but the energy difference $\Delta E_c$ between a monomer sitting on a favorable site (valley of the herringbone reconstruction) versus an unfavorable site (soliton wall). This corrugation energy scale is smaller but still significant, typically estimated to be $50 - 100$ meV, as observed in molecular assembly studies [43].

**(ii) The energy gain for an entire chain:** For a polymer chain containing $n$ monomers in the chain, the total energy gain, assuming that the entire chain is located (adsorbed) within the valley and each monomer is adsorbed there (i.e., the bob size is equal to the monomer size), would be: $E_{\text{total}} = n\, \Delta E_c$. For P3HT chains used in our experiments with molecular weight $20,000 - 45,000$ g/mol, $n \sim 120 - 270$ monomers. Using a conservative estimate of $\Delta E_c \approx 50$ meV we would get $E_{\text{total}} \approx 6$ to $13.5$ eV. Accordingly, under the employed experimental conditions, polymers remain attached to the substrate permanently (they cannot detach or evaporate). However, given that the provided thermal energy made movements of individual segments (monomers) possible, these polymers can diffuse on the surface, given that they can overcome existing surface potential barriers on the corrugated surface of the herringbone pattern of Au(111).

**(iii) Comparison with thermal energy:** At room temperature, $k_B T$ is approximately 26 meV. The energy ratio with respect to the adsorption energy of the whole chain is staggering: $\frac{E_{\text{total}}}{k_B T} \approx 230$ to $520$, implying a significant driving force for the entire chain to adsorb in the valleys. However, the energy ratio with respect to the adsorption energy of an individual monomer is much lower $\left[\frac{\Delta E_c}{k_B T} \approx 2\right]$, implying the realistic possibility of a sequential detachment-attachment of monomers. Due to the connectivity of the monomers along the polymer chain, the sequential movement of monomers enables the diffusion of polymer chains on the corrugated gold surface.

**(iv) Implications for monomer motion:** This enormous energy difference between $E_{\text{total}}$ and $k_B T$ has a crucial consequence. Every single monomer along the chain experiences a strong, directional driving force to move toward the nearest low energy valley. The potential gradient pulling monomers toward the valleys is immense compared to random thermal fluctuations.

Because monomers are covalently linked along the backbone, moving a single monomer from a peak to a valley would require cooperative motion of neighboring segments, necessarily involving chain stretching, deformation, and overcoming local barriers like $\pi - \pi$ stacking. The chain can rearrange segment by segment. However, at room temperature, the total time scale for such rearrangements of all segments along the whole chain can take hours to days.

Thus, even though the thermodynamic driving force is enormous, at room temperature the chains may need long times to overcome energy barriers (like $\pi - \pi$ stacking), potentially existing in the conformations of as-deposited chains. This situation may correspond to the partially adapted state shown schematically in Figure S4 (c), where the chain has not yet reached its equilibrium conformation.

### (c) The role of thermal energy (100°C annealing)

Annealing at 100°C increases the thermal energy to approximately 32 meV. For segmental motion along the backbone of a P3HT chain, the activation energy $E_a$ is reported to be in the range of few tens of meV [1,44], i.e., comparable to thermal energy at 100°C. Thus, increasing the temperature from room temperature to 100°C enhances the rate of segmental movements. Once a chain segment gains sufficient thermal energy to overcome local barriers, it experiences an overwhelmingly strong thermodynamic gradient ($\Delta E_c$) pulling it toward the energy valleys. Due to the chemical bonds between neighboring monomers, the process is not a diffusive random search of individual monomers but rather a directed, downhill migration of the whole chain. Segments that transiently escape the local barrier are rapidly channeled into potential minima by the large energy gradient ($\Delta E_c$ of $50 - 100$ meV per monomer) and their connection to neighboring monomers. This "zippering" effect propagates along the chain, pulling neighboring monomers sequentially into the valley.

Thus, 100°C annealing does not provide enough energy for chains to escape the surface potential entirely, the strong thiophene–Au interaction ensures firm adsorption [39]. Instead, the supplied thermal energy is sufficient for monomers to overcome local energy barriers, allowing their migration into valleys, reducing the time scales from days to hours.

**(d) Confinement in Surface Potential Valleys**

(i) **The valley as a 1D channel:** Once the chains have fully migrated into valleys, they find themselves confined within narrow, elongated channels defined by the herringbone reconstruction. These valleys are approximately $3-5$ nm wide (as shown in Figure S2). In this confined geometry, the chain can no longer explore the full 2D plane. Instead, it is restricted to movements of adsorption blobs along a quasi-1D channel.

(ii) **Blobs in confined geometry:** For a polymer confined in a linear channel of width $D$, the chain can be viewed as a linear string of blobs of size $\xi \lesssim D$. Within each blob, the chain remains unperturbed and behaves as an ideal chain [42].

**(e) The role of High-temperature annealing: 200°C**

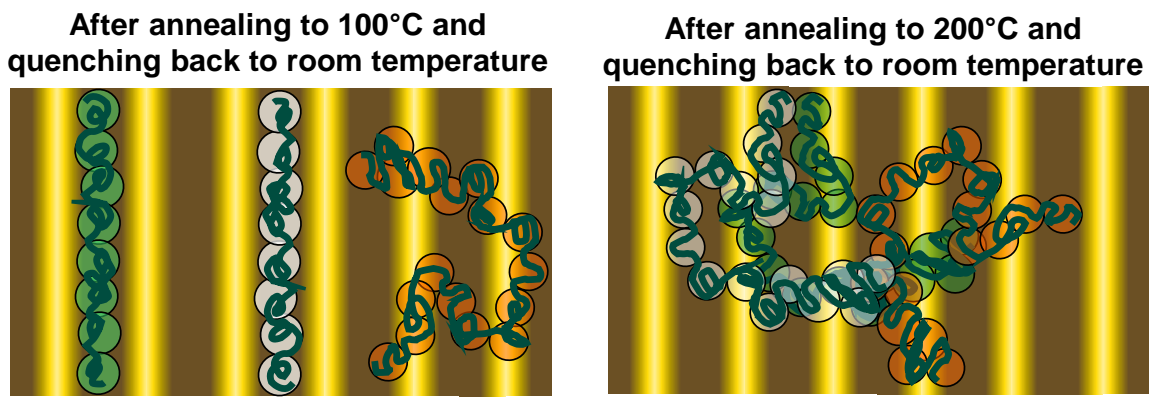

**Figure S5: Schematic representation of multi-chain clustering as a result of annealing to 200°C.**

At 200°C, $k_\mathrm{B}T \approx 41$ meV, polymer chains acquire a higher mobility, allowing polymer chains to diffuse over several nanometers. This mobility of polymer chains at elevated temperatures allowed for the encounter of multiple chains and the formation of multichain interactions via $\pi - \pi$ stacking among adjacent P3HT chains.

Furthermore, gold surface atoms also gain a higher surface mobility at 200°C. Studies of the dynamics of gold atoms at Au(111) surfaces and the resulting coarsening of surface structures consistently treat Au(111) as a "high mobility" surface where vacancies and adatoms can rearrange within the top layer well below the bulk melting point, implying that surface diffusion of Au atoms at 200 °C is significant, allowing for thermally induced reorganization along with possible disruptions in the reconstructed surface morphology[21,30,45].

This increased polymer mobility along with mobility of gold atoms on the substrate surface triggers a cascade of events leading to multi-chain clusters of polymer chains:

(i) **Coalescence of neighboring chains**: Chains previously confined in separate valleys are brought into close proximity by overcoming the soliton barriers.
(ii) **Polymer-polymer interactions dominate**: Once chains get in contact, strong $\pi - \pi$ interactions between P3HT backbones are capable of driving coalescence into multi-chain clusters. This represents a shift from a polymer-substrate to a polymer-polymer dominated assembly.
(iii) **Coarsening and size homogenization**: As more chains come into contact, they merge into larger clusters, driven by polymer diffusion and subsequent $\pi - \pi$ stacking.

Thus, 200°C annealing enable polymer diffusion and activates gold surface mobility, allowing polymer chains to diffuse over a distance of several nanometers allowing chains to come in contact, where $\pi - \pi$ interactions drive clustering.

# 5. Supplementary figures

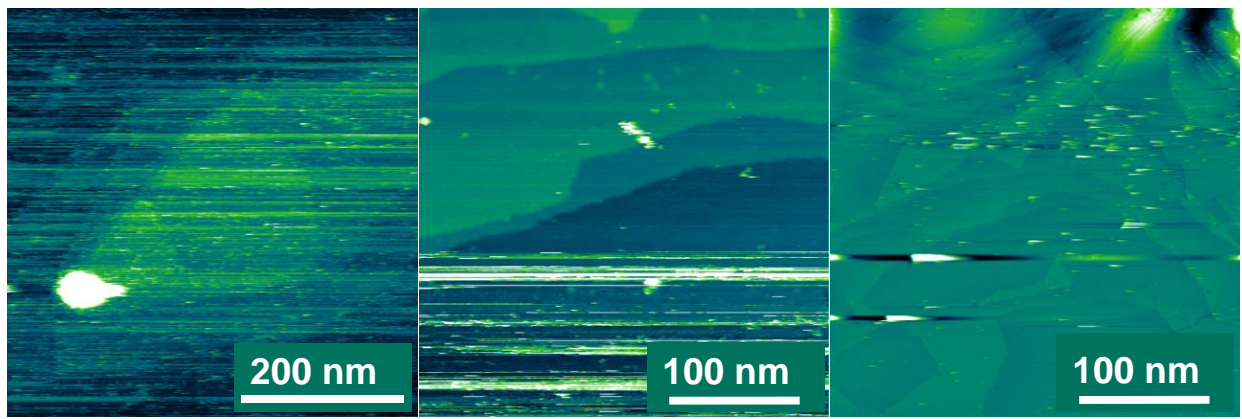

**Figure S6: STM topography images on electrosprayed P3HT on Au(R) before annealing to 100 °C, showing insufficient molecular resolution and providing no information on the deposited polymers.**
(a) Area: 500 × 500 nm², Bias: −2 V, Setpoint: 100 pA, (b) Area: 300 × 300 nm², Bias: −1.5 V, Setpoint: 400 pA, (c) Area: 300 × 300 nm², Bias: −0.5 V, Setpoint: 10 pA.

Figure S6 shows STM topography images of the P3HT/Au(111) sample acquired prior to annealing at 100 °C, where images consistently lacked molecular resolution and displayed strongly fluctuating contrast over the entire scanned area, independent of the chosen tunneling setpoint (tested between 10 pA and 400 pA). The absence of resolvable polymer chain features under all imaging conditions indicates that the tip–sample junction was not mechanically or electronically stable, which we attribute to the high lateral mobility of the as-deposited and weakly adsorbed P3HT chains at room temperature.

We expect that right after deposition under ultra-high vacuum conditions (vacuum can be considered as the worst solvent condition) polymer chains take up a globular state consisting of collapsed chain conformations, where intra-chain $\pi - \pi$ interactions may dominate. These collapsed chains can be thought as ball with minimum surface interactions and thus can be displaced rather easily by the movement of STM tip. Accordingly, for reliable high-resolution imaging, the polymer must be sufficiently immobilized on the Au(111) substrate. Experimentally, this anchoring was achieved by annealing the sample at 100 °C for 12 hours, followed by quenching to room temperature, which allowed the chains to settle in the valleys of the corrugations of the herringbone reconstruction of the Au(111) surface while overcoming the $\pi - \pi$ interaction within the collapsed conformations of the globular state. Thereby, the mobility of polymer chains was decreased which enabled their visualization by STM.

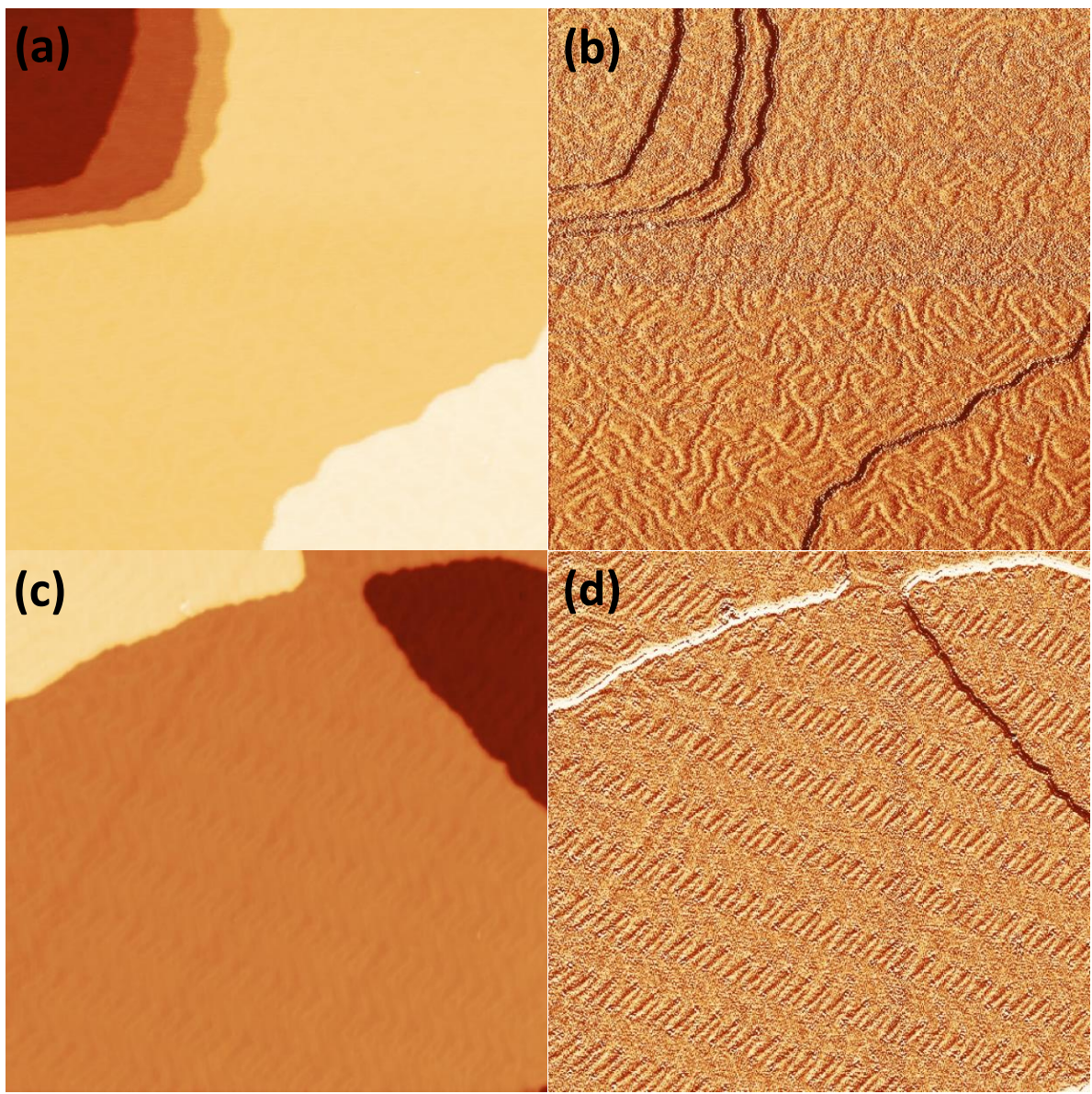

**Figure S7**: **Large-scale STM topography images of Au(111) with an irregular (IRR) and a regular (R) herringbone reconstruction.** (a, b) STM topography image and current map, respectively, of Au(IRR). (c, d) STM topography image and current map, respectively, of Au(R). (a, b) Area: 150 × 150 nm², Bias: -1.5 V, Setpoint: 3000 pA. (c, d) Area: 180 × 180 nm², Bias: −2.0 V, Setpoint: 2000 pA.

The large-scale STM images in Figure S7 depict the evolution of the Au(111) surface from a disordered state to a well-ordered herringbone reconstruction following the cleaning procedure and quenching to room temperature. Images acquired within 24 hours after the end of the cleaning procedure (Figure S7 (a), topography; Figure S7 (b), current map) show extended Au(111) terraces. However, the corresponding current map reveals an irregular, spatially disordered herringbone pattern, indicating that the reconstruction was not yet equilibrated and lacked long-range order. By contrast, after the sample was held at room temperature under UHV for approximately 68 hours, large-area STM topography images and current maps (Figures S7 (c) and S7 (d), respectively) revealed wide terraces featuring a regular, periodic herringbone reconstruction that extended over hundreds of nanometers.

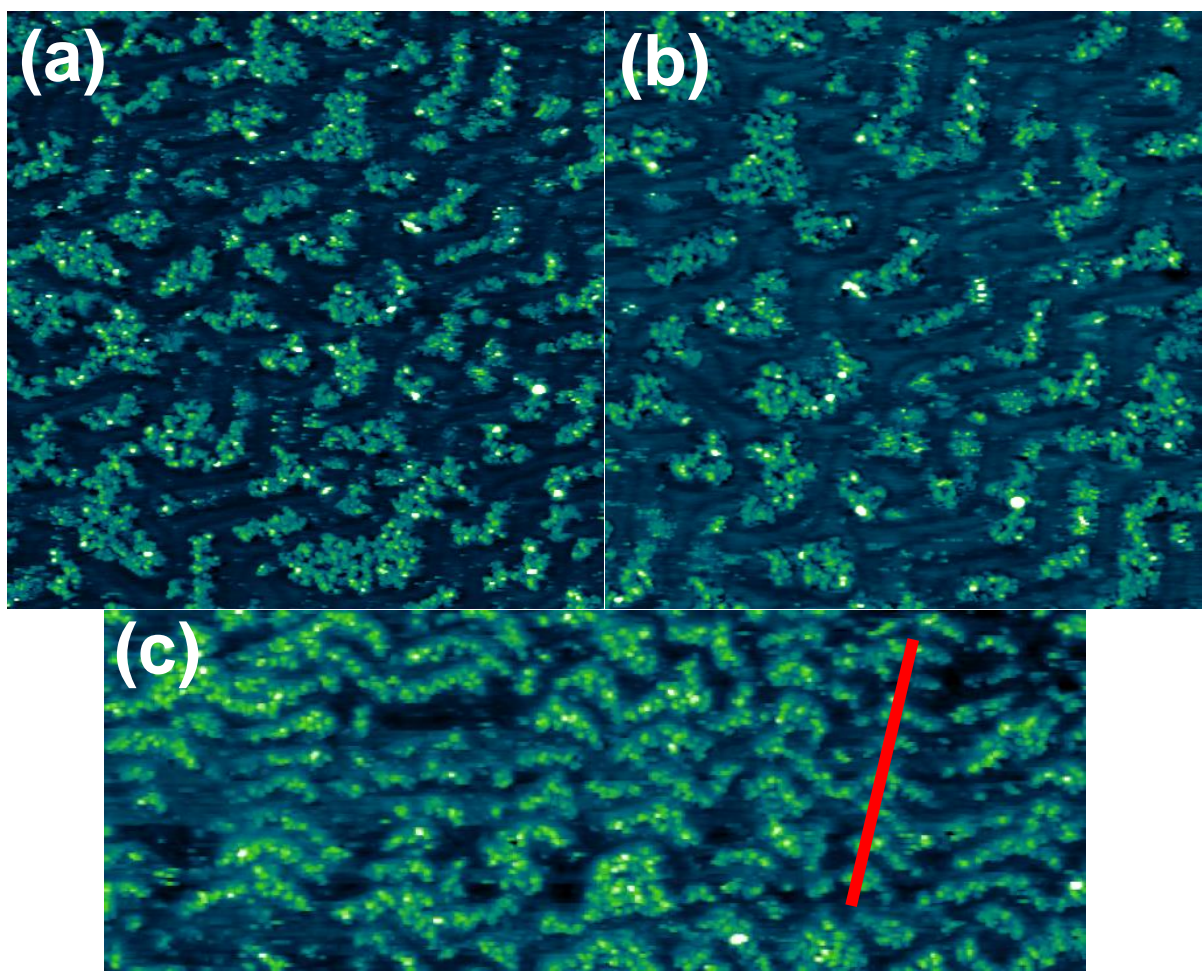

**Figure S8**: **Large-scale STM topography image showing that P3HT chains replicated the herringbone reconstruction periodically**. (a) Area: $82 \times 82$ nm$^2$, Bias: $-2.0$ V, Setpoint: 10 pA. (b) Area: $60 \times 60$ nm$^2$, Bias: $-2.0$ V, Setpoint: 10 pA. (c) Area: $169 \times 62$ nm$^2$, Bias: $-2.55$ V, Setpoint: 10 pA.

Figure S8 presents large-scale STM topography images of P3HT clusters recorded at multiple, spatially separated regions of the Au(111) surface, demonstrating that the polymer conformations consistently replicated the underlying herringbone reconstruction over hundreds of nanometers. In each image, P3HT polymer chains aligned along the directions defined by the herringbone soliton walls and elbows, indicating that the substrate reconstruction acted as a long-range template for chain clusters. From these large-scale images such as Figure S8 (c), a periodic distance of $d_\mathrm{P} \approx 7$ nm between polymer chains can be estimated along the red line. This $d_\mathrm{P} \approx 7$ nm period is comparable to the distance between consecutive fcc sites as shown in Figure S2 (a). The observation of the same morphology and orientation of patterns measured at different positions on the sample, acquired in independent scans and through several imaging sessions, confirmed that the structures established by the P3HT chains were stable on the experimental timescale and that the morphology of the adsorbed P3HT chains reported in the main text was highly reproducible and representative of the global film morphology rather than a rare, site-specific arrangement of P3HT chains.

In an attempt to quantitatively validate the monomer footprint, Figure S9 correlates monomer-resolved STM measurements of P3HT chains with molecular models. Figure S9 (a) and (b) show small-scale STM topography images acquired at room temperature and the corresponding derivative map, where individual thiophene units along two intersecting P3HT chains are clearly resolved. From the real-space image contrast, we were able to estimate the projected area occupied by a single monomer ($A_\mathrm{m}$). To crosscheck these experimental values, an energy-minimized structure of an isolated P3HT monomer was generated in Avogadro using the MMFF94 force field, and its lateral molecular footprint was calculated, yielding a value of $A_\mathrm{m}$ that closely matched the STM-derived estimate within the experimental uncertainty, as summarized in the Table in Figure S9 (e). Similarly, the monomer–monomer spacing ($d_\mathrm{mm}$) extracted from the STM images in Figure S9 (a) and (b) was compared to the inter-monomer separation obtained from a dimer model of two covalently linked P3HT units (Figure S9 (d)). This shows an excellent agreement between the experimentally measured repeat distance and that predicted by the molecular model (Table in Figure S9 (e)), thereby confirming that the observed periodic contrast in STM images indeed reflects the intrinsic monomeric repeat unit along the P3HT backbone.

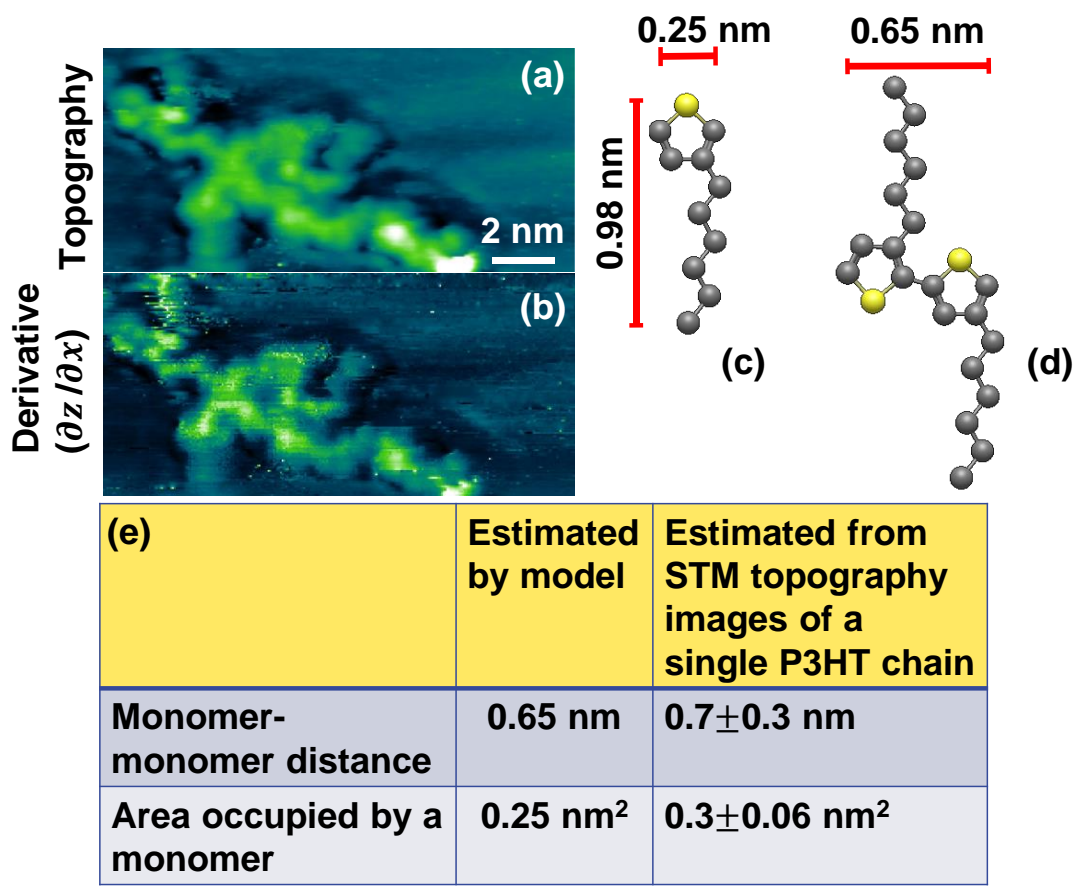

**Figure S9: Estimation and comparison of the monomer–monomer distance ($d_{mm}$) and the area occupied by a single monomer ($A_m$) obtained from molecular modelling and STM topography images.** (a) STM topography image showing P3HT polymer chains with monomeric resolution that intersect at a point. (b) Corresponding derivative image of (a) showing enhanced monomer contrast. (c) Optimized model of a single P3HT monomer generated in Avogadro using the MMFF94 (Merck Molecular Force Field 94). (d) Model of two covalently linked monomers showing an inter-monomer separation of 0.65 nm. (e) Table comparing $d_{mm}$ and $A_m$ values extracted from the molecular model and from STM topographs. Error bars denote the standard deviation of measurements. Scan area: 15 × 10 nm²; setpoint: 10 pA; sample bias: −2.0 V.

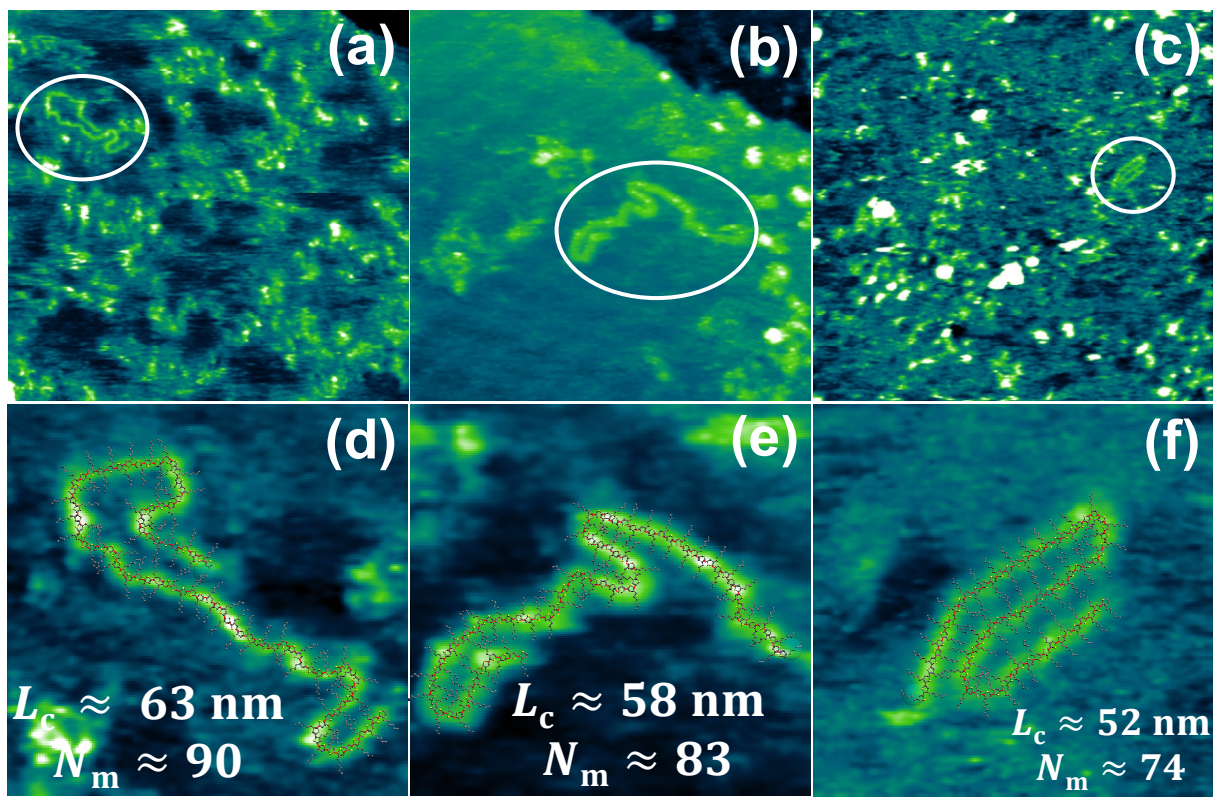

**Figure S10**: **Electrospray deposition of P3HT on a gold surface with an irregular corrugation pattern (Au(IRR)) showing different conformations of individual polymer chains.** (a, b, c) Large-scale STM topography images with selected polymer chains highlighted by white circles. (d, e, f) Small-scale STM topography images of the individual polymer chains in the regions highlighted in (a, b, c) with white circles, respectively. Images in (d, e, f) were superimposed with the periodic arrangement (0.7 nm along the backbone) of a model of a monomer (Figure S9 (c)). (a) Scanned area: $60 \times 60$ nm$^2$ at bias: $-1.5$ V, and setpoint: 5 pA. (b) Scanned area: $40 \times 40$ nm$^2$ at bias: $-1.53$ V, setpoint: 5 pA. (c) Scanned area: $47 \times 47$ nm$^2$ at bias: $-1.8$ V, setpoint: 9 pA. (d) Scanned area: $15 \times 15$ nm$^2$ at bias: $-1.5$ V, and setpoint: 5 pA. (e) Scanned area: $15 \times 15$ nm$^2$ at bias: $-1.53$ V, setpoint: 5 pA. (f) Scanned area: $15 \times 15$ nm$^2$ at bias: $-1.8$ V, setpoint: 9 pA. $L_c$ is the contour length, $N_m$ is the number of monomers. Model of a single P3HT monomer generated in Avogadro using the MMFF94 (Merck Molecular Force Field 94) force field was in combination with LMAPper software to superimpose the periodic arrangement of monomers on the STM images.

Figures S10 (a–c) present large-area STM topography images of P3HT deposited by electrospraying on Au(111) with an irregular herringbone reconstruction (Au(IRR)), acquired after annealing at 100 °C. In contrast to the extended, herringbone-aligned clusters observed on regularly reconstructed Au(R) (Figure S9), the polymer morphology on Au(IRR) was qualitatively different. Across all three images acquired from widely separated regions of the same sample, the chains exhibited no obvious long-range periodic arrangement or directional alignment replicating the underlying surface. The consistent absence of surface-guided templating across multiple, independent regions demonstrated that the irregular adsorption behavior of chains on Au(IRR) is a characteristic feature and not a defect-specific anomaly.

Nevertheless, a small number of individual chains adopted well-defined random-walk-like conformations, as highlighted by white circles in Figures S10 (a–c). Higher-resolution STM images of these selected chains are shown in Figures S10 (d–f). To validate the relation between the observed contrast and molecular features of P3HT chains, these high-resolution images were superimposed with a periodic array of P3HT monomer models (placed at a spacing of 0.7 nm), generated using the MMFF94 force field in Avogadro and manually aligned using LMAPper software. The excellent agreement between the model and the contrast in the STM images confirmed that the resolved features corresponded to individual thiophene repeat units along the polymer backbone, with contour length $L_c$ and number $N_m$ of monomers, as indicated for each chain.

To understand the origin of the two conformational motifs observed in Figure S10, Figure S11 (a) shows the bare Au(IRR) surface before deposition. The STM topography image revealed a disordered corrugation pattern lacking a well-defined periodic length scale. A schematic overlay of representative chain conformations is included to provide a possible interpretation of the two adsorption motifs observed experimentally.

Figures S11 (b) and S11 (c) present higher-resolution STM images of P3HT chains corresponding to these motifs. In Figure S11 (b), the polymer chain followed the local direction of the irregular corrugation pattern, adapting its trajectory to the underlying topography. In Figure S11 (c), the chain adopted a collapsed conformation, confined within a particularly favorable low-energy region of the disordered potential landscape. These images correspond to the blue circle (corrugation-following) and red circle (collapsed) schematic overlays in Figure S11 (a), respectively.

Taken together, Figures S10 and S11 demonstrate that on Au(IRR), where the surface potential lacks a regular periodic structure, the polymer population is dominated by non-templated conformations. The occasional observation of locally ordered or collapsed chains reflects the influence of deep potential minima in an otherwise random landscape. This behavior contrasts sharply with the uniform, corrugation-following clusters formed on regularly reconstructed Au(R), underscoring the critical role of a well-defined surface periodicity in directing polymer organization.

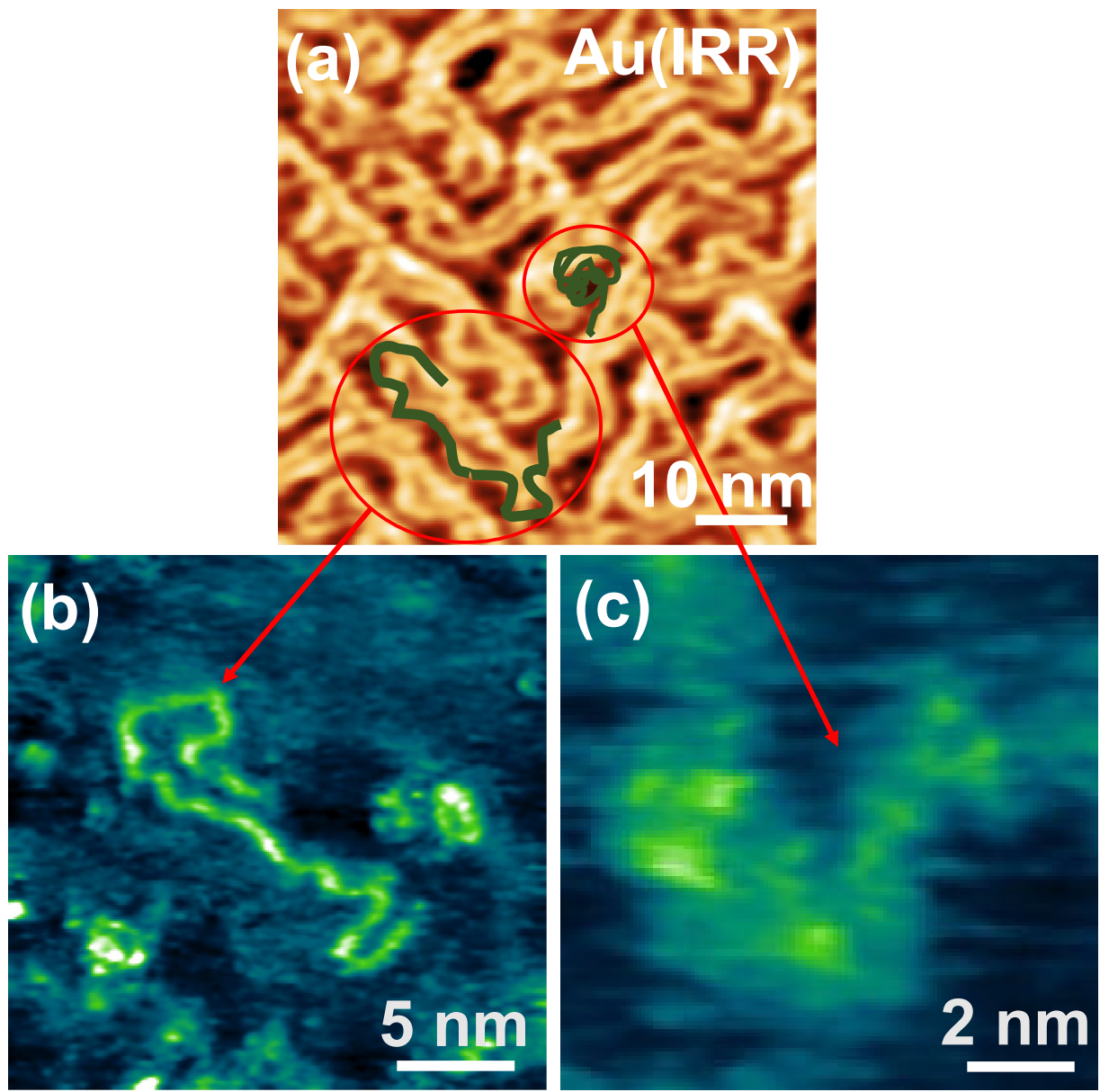

**Figure S11**: (a) STM topography image of Au (111) with irregular surface reconstruction before electrospray deposition, including the overlay of schematic of polymer chains with different conformations. (b, c) STM images taken after electrospray deposition of P3HT, corresponding to the schematic representations shown in (a). (a) Scanned area: $60 \times 60$ nm$^2$ at bias: $-1.5$ V, and setpoint: 300 pA. (b) Scanned area: $23 \times 23$ nm$^2$ at bias: $-1.51$ V, setpoint: 5 pA. (c) Scanned area: $10 \times 10$ nm$^2$ at bias: $-1.51$ V, setpoint: 5 pA.

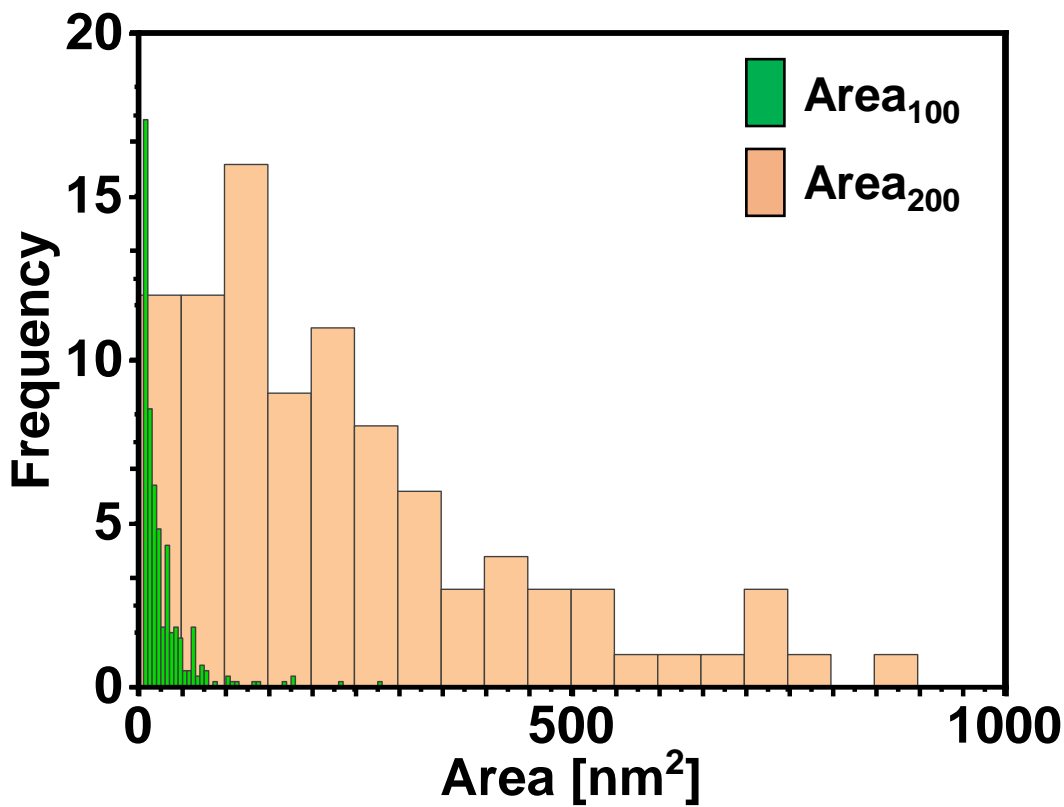

**Figure S12:** Histograms showing the distribution of cluster areas observed in STM topography after annealing at 100 °C (Green color (Area$_{100}$)) and 200 °C (Orange color (Area$_{200}$)).

Figure S12 shows histogram plots of the areas covered by each cluster, estimated from a large set of STM images for the two series of experiments (i.e., samples after annealing at 100 °C (Area$_{100}$), and samples after annealing at 200 °C (Area$_{200}$)). A significant increase in the average area per cluster was observed after annealing at 200°C compared to samples annealed at 100 °C indicates that the number of polymers per cluster has grown significantly. Furthermore, after annealing at 200 °C, the total area covered by all clusters decreased compared to annealing to 100 °C. The ratio ($\frac{\phi_{200}}{\phi_{100}}$) of the area fraction $\phi_{iii}$ with being defined as $\phi_{iii} = \frac{\text{total area covered by all clusters (at temperature } T=iii \text{ °C)}}{\text{total scanned area}}$ ) was estimated to be $\frac{\phi_{200}}{\phi_{100}} \approx 0.5$. This decrease in area suggests that the clusters became thicker, consistent with superposition and stacking of chains. Concurrently, the number density of well-separated clusters per unit area (i.e., $n_{100}$ and $n_{200}$) decreased, yielding $\frac{n_{200}}{n_{100}} \approx 0.05$. The corresponding mean area $\overline{\text{Area}}_{iii}$ of a cluster increased, yielding a ratio of $\frac{\overline{\text{Area}}_{200}}{\overline{\text{Area}}_{100}} \approx 10$, suggesting lateral clustering of polymer chains. Together, these observations demonstrate that heating to 200 °C promoted both vertical stacking and lateral clustering of polymer chains.

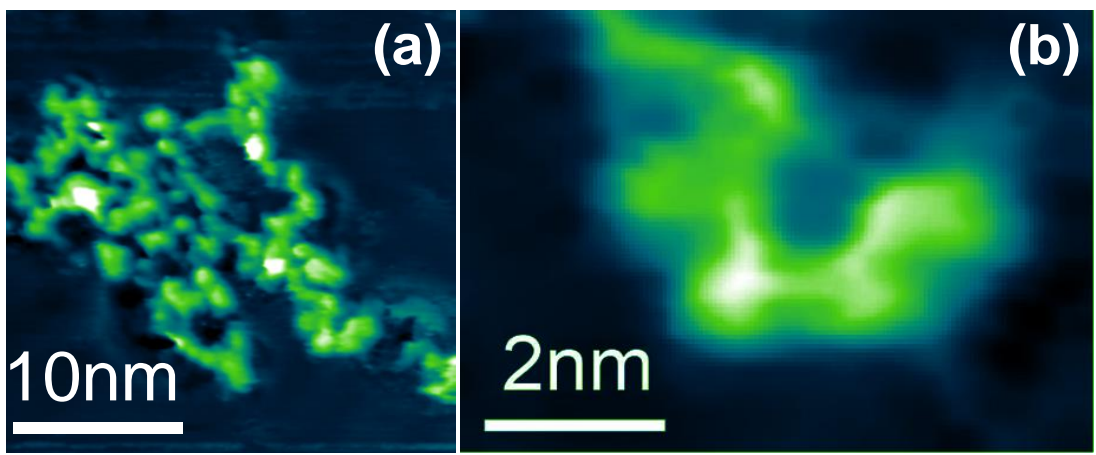

**Figure S13:** High-resolution STM topography image of large clusters observed after 200 °C annealing, showing interconnected blobs. (a) Large-scale image of a cluster of polymer chains. (b) Small-scale image of selected blobs observed in (a). The size of image (a) is $27 \times 27$ nm² measured at a setpoint of 60 pA, and a bias voltage of $-1.7$ V. The size of image (b) is $7 \times 5$ nm² measured at a setpoint of 60 pA, and a bias voltage of $-1.7$ V.

Figure S13 presents a high-resolution STM topography image of a representative cluster of P3HT chains, selected from those depicted in Figure 4 of the main text. The image reveals that the clusters consist of interconnected domains, each with an approximate diameter of $2 - 3$ nm. This characteristic length scale can be linked to the adsorption blob discussed in the context of Figure S4 and Figure 1 of main text. The formation of such domains indicates that within each blob, the polymer segments behave as an ideal chain, with chain–substrate interactions becoming significant only beyond the $2 - 3$ nm scale.

# References

1. Bhatta, R. S., Yimer, Y. Y., Tsige, M. & Perry, D. S. Conformations and torsional potentials of poly(3-hexylthiophene) oligomers: Density functional calculations up to the dodecamer. *Computational and Theoretical Chemistry* **995**, 36–42 (2012).
2. Khlaifia, D., Chemek, M. & Alimi, K. DFT/TDDFT approach: an incredible success story in prediction of organic materials properties for photovoltaic application. (2020).
3. Thorley, K. J. & Nielsen, C. B. Conformational Analysis of Conjugated Organic Materials: What Are My Heteroatoms Really Doing? *ChemPlusChem* **89**, e202300773 (2024).
4. Zhugayevych, A., Mazaleva, O., Naumov, A. & Tretiak, S. Lowest-Energy Crystalline Polymorphs of P3HT. *J. Phys. Chem. C* **122**, 9141–9151 (2018).
5. Casalegno, M., Famulari, A. & Meille, S. V. Modeling of Poly(3-hexylthiophene) and Its Oligomer's Structure and Thermal Behavior with Different Force Fields: Insights into the Phase Transitions of Semiconducting Polymers. *Macromolecules* **55**, 2398–2412 (2022).
6. Raithel, D., Simine, L., Pickel, S., Schötz, K., Panzer, F., Baderschneider, S., Schiefer, D., Lohwasser, R., Köhler, J., Thelakkat, M., Sommer, M., Köhler, A., Rossky, P. J. & Hildner, R. Direct observation of backbone planarization via side-chain alignment in single bulky-substituted polythiophenes. *Proc. Natl. Acad. Sci. U.S.A.* **115**, 2699–2704 (2018).
7. Yan, X., Xiong, M., Deng, X.-Y., Liu, K.-K., Li, J.-T., Wang, X.-Q., Zhang, S., Prine, N., Zhang, Z., Huang, W., Wang, Y., Wang, J.-Y., Gu, X., So, S. K., Zhu, J. & Lei, T. Approaching disorder-tolerant semiconducting polymers. *Nat Commun* **12**, 5723 (2021).
8. Sabury, S., Jones, A. L., Schopp, N., Nanayakkara, S., Chaney, T. P., Coropceanu, V., Marder, S. R., Toney, M. F., Brédas, J.-L., Nguyen, T.-Q. & Reynolds, J. R. Manipulating Backbone Planarity of Ester Functionalized Conjugated Polymer Constitutional Isomer Derivatives Blended with Molecular Acceptors for Controlling Photovoltaic Properties. *Chem. Mater.* **36**, 11656–11668 (2024).
9. Sutton, C., Körzdörfer, T., Gray, M. T., Brunsfeld, M., Parrish, R. M., Sherrill, C. D., Sears, J. S. & Brédas, J.-L. Accurate description of torsion potentials in conjugated polymers using density functionals with reduced self-interaction error. *The Journal of Chemical Physics* **140**, 054310 (2014).
10. Baggioli, A. & Famulari, A. On the inter-ring torsion potential of regioregular P3HT: a first principles reexamination with explicit side chains. *Phys. Chem. Chem. Phys.* **16**, 3983 (2014).
11. Torras, J. Building a Torsional Potential between Thiophene Rings to Illustrate the Basics of Molecular Modeling. *J. Chem. Educ.* **100**, 395–401 (2023).

12. Perkins, M. A., Cline, L. M. & Tschumper, G. S. Torsional Profiles of Thiophene and Furan Oligomers: Probing the Effects of Heterogeneity and Chain Length. *J. Phys. Chem. A* **125**, 6228–6237 (2021).

13. Pantawane, S. & Gekle, S. Temperature-Dependent Conformation Behavior of Isolated Poly(3-hexylthiopene) Chains. *Polymers* **14**, 550 (2022).

14. Mackie, I. D., McClure, S. A. & DiLabio, G. A. Binding in Thiophene and Benzothiophene Dimers Investigated By Density Functional Theory with Dispersion-Correcting Potentials. *J. Phys. Chem. A* **113**, 5476–5484 (2009).

15. Tsuzuki, S., Honda, K. & Azumi, R. Model Chemistry Calculations of Thiophene Dimer Interactions: Origin of π-Stacking. *J. Am. Chem. Soc.* **124**, 12200–12209 (2002).

16. Gupta, S., Chatterjee, S., Zolnierczuk, P., Nesterov, E. E. & Schneider, G. J. Impact of Local Stiffness on Entropy Driven Microscopic Dynamics of Polythiophene. *Sci Rep* **10**, 9966 (2020).

17. Liang, J., Ouyang, X. & Cao, Y. Interfacial and confined molecular-assembly of poly(3-hexylthiophene) and its application in organic electronic devices. *Science and Technology of Advanced Materials* **23**, 619–632 (2022).

18. Förster, S., Kohl, E., Ivanov, M., Gross, J., Widdra, W. & Janke, W. Polymer adsorption on reconstructed Au(001): A statistical description of P3HT by scanning tunneling microscopy and coarse-grained Monte Carlo simulations. *The Journal of Chemical Physics* **141**, 164701 (2014).

19. Greco, C., Melnyk, A., Kremer, K., Andrienko, D. & Daoulas, K. Ch. Generic Model for Lamellar Self-Assembly in Conjugated Polymers: Linking Mesoscopic Morphology and Charge Transport in P3HT. *Macromolecules* **52**, 968–981 (2019).

20. Harten, U., Lahee, A. M., Toennies, J. P. & Wöll, Ch. Observation of a Soliton Reconstruction of Au(111) by High-Resolution Helium-Atom Diffraction. *Phys. Rev. Lett.* **54**, 2619–2622 (1985).

21. Barth, J. V., Brune, H., Ertl, G. & Behm, R. J. Scanning tunneling microscopy observations on the reconstructed Au(111) surface: Atomic structure, long-range superstructure, rotational domains, and surface defects. *Phys. Rev. B* **42**, 9307–9318 (1990).

22. Carter, C. B. & Hwang, R. Q. Dislocations and the reconstruction of (111) fcc metal surfaces. *Phys. Rev. B* **51**, 4730–4733 (1995).

23. Bartelt, N. C. & Thürmer, K. Structure and energetics of the elbows in the Au(111) herringbone reconstruction. *Phys. Rev. B* **104**, (2021).

24. Hinaut, A., Scherb, S., Yao, X., Liu, Z., Song, Y., Moser, L., Marot, L., Müllen, K., Glatzel, T., Narita, A. & Meyer, E. Stable Au(111) Hexagonal Reconstruction Induced by Perchlorinated Nanographene Molecules. *J. Phys. Chem. C* **128**, 18894–18900 (2024).

25. Bulou, H. & Massobrio, C. New Atomic Mechanism of Preferential Nucleation on the Herringbone Reconstruction of Au(111). *J. Phys. Chem. C* **112**, 8743–8746 (2008).
26. Ripani, G., Flachmüller, A., Peter, C. & Palleschi, A. Coarse-Grained Simulation of the Adsorption of Water on Au(111) Surfaces Using a Modified Stillinger–Weber Potential. *ACS Omega* **5**, 31055–31059 (2020).
27. Casari, C. S., Foglio, S., Siviero, F., Li Bassi, A., Passoni, M. & Bottani, C. E. Direct observation of the basic mechanisms of Pd island nucleation on Au(111). *Phys. Rev. B* **79**, (2009).
28. Böhringer, M., Morgenstern, K., Schneider, W.-D., Berndt, R., Mauri, F., De Vita, A. & Car, R. Two-Dimensional Self-Assembly of Supramolecular Clusters and Chains. *Phys. Rev. Lett.* **83**, 324–327 (1999).
29. Clair, S., Pons, S., Brune, H., Kern, K. & Barth, J. V. Mesoscopic Metallosupramolecular Texturing by Hierarchic Assembly. *Angew Chem Int Ed* **44**, 7294–7297 (2005).
30. Spurgeon, P. M., Lai, K. C., Han, Y., Evans, J. W. & Thiel, P. A. Fundamentals of Au(111) Surface Dynamics: Coarsening of Two-Dimensional Au Islands. *J. Phys. Chem. C* **124**, 7492–7499 (2020).
31. Li, P. & Ding, F. Origin of the herringbone reconstruction of Au(111) surface at the atomic scale. *Sci. Adv.* **8**, eabq2900 (2022).
32. Repain, V., Berroir, J. M., Rousset, S. & Lecoeur, J. Reconstruction, step edges and self-organization on the Auž111/ surface. (2000).
33. Kowalczyk, P., Kozlowski, W., Klusek, Z., Olejniczak, W. & Datta, P. K. STM studies of the reconstructed Au(111) thin-film at elevated temperatures. *Applied Surface Science* **253**, 4715–4720 (2007).
34. Noh, J., Ito, E., Nakajima, K., Kim, J., Lee, H. & Hara, M. High-Resolution STM and XPS Studies of Thiophene Self-Assembled Monolayers on Au(111). *J. Phys. Chem. B* **106**, 7139–7141 (2002).
35. Tachibana, M., Yoshizawa, K., Ogawa, A., Fujimoto, H. & Hoffmann, R. Sulfur–Gold Orbital Interactions which Determine the Structure of Alkanethiolate/Au(111) Self-Assembled Monolayer Systems. *J. Phys. Chem. B* **106**, 12727–12736 (2002).
36. Liu, Y.-F., Krug, K. & Lee, Y.-L. Self-organization of two-dimensional poly(3-hexylthiophene) crystals on Au(111) surfaces. *Nanoscale* **5**, 7936 (2013).
37. Scifo, L., Dubois, M., Brun, M., Rannou, P., Latil, S., Rubio, A. & Grévin, B. Probing the Electronic Properties of Self-Organized Poly(3-dodecylthiophene) Monolayers by Two-Dimensional Scanning Tunneling Spectroscopy Imaging at the Single Chain Scale. *Nano Lett.* **6**, 1711–1718 (2006).

38. Rodriguez, J. A., Dvorak, J., Jirsak, T., Liu, G., Hrbek, J., Aray, Y. & González, C. Coverage Effects and the Nature of the Metal−Sulfur Bond in S/Au(111): High-Resolution Photoemission and Density-Functional Studies. *J. Am. Chem. Soc.* **125**, 276–285 (2003).
39. Tonigold, K. & Groß, A. Adsorption of small aromatic molecules on the (111) surfaces of noble metals: A density functional theory study with semiempirical corrections for dispersion effects. *The Journal of Chemical Physics* **132**, 224701 (2010).
40. Yamamoto, K., Ochiai, S., Wang, X., Uchida, Y., Kojima, K., Ohashi, A. & Mizutani, T. Evaluation of molecular orientation and alignment of poly(3-hexylthiophene) on Au (111) and on poly(4-vinylphenol) surfaces. *Thin Solid Films* **516**, 2695–2699 (2008).
41. Adhikari, S., Tang, H., Neupane, B., Ruzsinszky, A. & Csonka, G. I. Molecule-surface interaction from van der Waals-corrected semilocal density functionals: The example of thiophene on transition-metal surfaces. *Phys. Rev. Materials* **4**, 025005 (2020).
42. Rubinstein, M. & Colby, R. H. *Polymer Physics*. (Oxford University Press, Oxford, 2023).
43. Takeuchi, N., Chan, C. T. & Ho, K. M. Au(111): A theoretical study of the surface reconstruction and the surface electronic structure. *Phys. Rev. B* **43**, 13899–13906 (1991).
44. Kanai, K., Miyazaki, T., Suzuki, H., Inaba, M., Ouchi, Y. & Seki, K. Effect of annealing on the electronic structure of poly(3-hexylthiophene) thin film. *Physical Chemistry Chemical Physics* **12**, 273–282 (2010).
45. Baber, A. E., Jensen, S. C., Iski, E. V. & Sykes, E. C. H. Extraordinary Atomic Mobility of Au{111} at 80 Kelvin: Effect of Styrene Adsorption. *J. Am. Chem. Soc.* **128**, 15384–15385 (2006).